\documentclass[reprint,onecolumn,showpacs,nofootinbib,prd,amssymb]{revtex4}

\usepackage{latexsym}
\usepackage{graphicx,epsf, epsfig, amssymb, amsmath}% Include figure files

\usepackage{color}
\usepackage{relsize}
\usepackage{textcomp}
\usepackage{float}
\newcommand{\pd}[2]{\frac{\partial #1}{\partial #2}} 

\newcommand{\nablaperp}{\mathbin{^{\mathsmaller{\perp}}\mkern-3mu \nabla}}

\newcommand{\Fperp}{\mathbin{^{\mathsmaller{\perp}}\mkern-3mu F}}

\usepackage[normalem]{ulem}

\begin{document}

%%%%%%%%%%%%%%%%%%%%%%%%%%%%%%%%%%%%%%%%%%%%%%%%%%%%%%%%%%%%%%%%%%%%%%%%%%%%%%
\title{Dissipative relativistic magnetohydrodynamics of a multicomponent mixture and its application to neutron stars}
%%%%%%%%%%%%%%%%%%%%%%%%%%%%%%%%%%%%%%%%%%%%%%%%%%%%%%%%%%%%%%%%%%%%%%%%%%%%%%
\author{V.~A.~Dommes}
\email[e-mail: ]{vasdommes@gmail.com}
\author{M.~E.~Gusakov}
\author{P.~S.~Shternin}
\affiliation{Ioffe Institute,
Politekhnicheskaya 26, 194021 St.~Petersburg, Russia
}

%%%%%%%%%%%%%%%%%%%%%%%%%%%%%%%%%%%%%%%%%%
\begin{abstract} 
We formulate hydrodynamic equations for nonsuperfluid multicomponent magnetized charged relativistic mixtures,
taking into account chemical reactions as well as viscosity, diffusion, thermodiffusion, and thermal conductivity effects.
The resulting equations have a rather simple form and can be readily applied,
e.g., for studying magnetothermal evolution of neutron stars.
We also establish a link between our formalism and the results known in the literature,
and express the phenomenological diffusion coefficients
through momentum transfer rates which are calculated from microscopic theory.
\end{abstract}
%%%%%%%%%%%%%%%%%%%%%%%%%%%%%%%%%%%%%%%%%%

\date{\today}

%47.32.C- 	Vortex dynamics
%
%47.32.-y 	Vortex dynamics; rotating fluids (for vortices in superfluid helium, see 67.25.dk and 67.30.he)
%
%47.37.+q 	Hydrodynamic aspects of superfluidity; quantum fluids (for transport and hydrodynamics of normal and superfluid phase of 4He, see 67.25.bf, and 67.25.dg, respectively; for transport and hydrodynamics of normal and superfluid phase of 3He, see 67.30.eh, and 67.30.hb, respectively)
%
%04. 	General relativity and gravitation (for astrophysical aspects, see 95.30.Sf Relativity and gravitation; for relativistic aspects of cosmology, see 98.80.Jk; for special relativity, see 03.30.+p)
%
%03.30.+p 	Special relativity
%
%04.40.Nr 	Einstein-Maxwell spacetimes, spacetimes with fluids, radiation or classical fields
%
%04.40.Dg 	Relativistic stars: structure, stability, and oscillations (see also 97.60.-s Late stages of stellar evolution)
%
%95.30.Lz	Hydrodynamics
% 
%95.30.Qd	Magnetohydrodynamics and plasmas (see also 52.30.Cv and 52.72.+v—in physics of plasmas)
% 
%95.30.Sf	Relativity and gravitation (see also section 04 General relativity and gravitation; 98.80.Jk Mathematical and relativistic aspects of cosmology)
% 
%95.30.Tg	Thermodynamic processes, conduction, convection, equations of state
% 97.60.Jd Neutron stars (see also 26.60.-c Nuclear matter aspects of neutron stars in—Nuclear physics) 

%PACS
%(1) Choose no more than four index number codes.
%(2) Place your principal index code first.
%(3) Always choose the lowest-level code available.
%(4) Always include the check characters.

%%%%%%%%%%%%%%%%%%%%%%%%%%%%%%%%%%%%%%

\maketitle

%%%%%%%%%%%%%%%%%%%%%%%%%%%%%%%%%%%%%%%%%%%%%%%%%%%%%%%%%%%%%%%%%%%%
\section{Introduction}
\label{sec:intro}
%%%%%%%%%%%%%%%%%%%%%%%%%%%%%%%%%%%%%%%%%%%%%%%%%%%%%%%%%%%%%%%%%%%%

Observations of neutron stars (NSs) 
contain a wealth of potentially important information 
about the properties of superdense matter in their interiors \cite{watts_etal16,kaspi10,haskell15,hm15,ppp15,of16,kb17}.
In order to extract this information, however,
one has to build realistic models, 
allowing the theoretical study of the NS dynamics.
Such models should account for 
various particle species in the core (neutrons, protons, electrons with an admixture
of muons, and, possibly, hyperons and/or quarks), magnetic field,
baryon superfluidity, and effects of the general theory of relativity.
Clearly, the construction of such models is a complex theoretical problem, which is under intensive development now
(e.g., Refs. \cite{gas11,vigano_etal13,gagl15,awv16,andersson12,ach17,ahdc17,adhc17,gd16,gko17,gusakov19}).

For example, in studies of magnetothermal evolution,
one has to account for the fact that the magnetic field in
superconducting NS cores can be confined to Abrikosov vortices.
Then the problem of magnetic field evolution
reduces to the analysis of motion of vortices
under the action of various forces exerted on them by different particle species
(neutrons, protons, electrons, muons, etc.),
which move with different velocities and interact with one another.
Smooth-averaged relativistic magnetohydrodynamic (MHD) equations,
suitable for describing the evolution of such a system at finite temperatures,
were formulated
in Ref.~\cite{gd16}, neglecting diffusion of normal (nonsuperfluid and nonsuperconducting) particles
(see also a number of related works \cite{gas11,gusakov16,awv16} in this direction).
However, diffusion is known to play an important role,
affecting not only dissipation of the magnetic field, but also its nondissipative evolutionary timescales
\cite{gr92,su95,blb18,crv17,eprgv16,gagl15,papm17,gko17,og18,dg17,pap17}.

Therefore, initially, our main goal was to generalize the equations obtained in Ref.~\cite{gd16} to
allow for diffusion.
During this work, it quickly became clear
that even in the absence of baryon superfluidity and superconductivity there are certain gaps in the literature 
devoted to magnetized relativistic mixtures,
which become especially apparent in the NS context.
First of all, a majority of works on relativistic dissipative MHD
(see, e.g., Refs.~\cite{lichnerowicz67,bo78,komissarov07,plrr09})
postulate a simplified version of Ohm's law, which 
includes the electric field but 
ignores gradients of thermodynamic functions (chemical potentials and temperature).
The latter terms are generally present even in the single-fluid MHD, 
not to mention its multifluid extensions
\cite{brag65,vanErklens1977PhyA-2}, and can be important for NS conditions \cite{gr92,su95}.
There are only a few papers in which the generalized Ohm's law for relativistic MHD
is derived simultaneously with the basic dynamic equations
(see, e.g., Ref. \cite{kt08} and a recent series of papers by Andersson \textit{et al.} 
\cite{andersson12,ach17,adhc17,ahdc17}).
In our opinion, the main shortcoming of these works (if we talk about applications to neutron stars)
is their excessive complexity and lack of transparency of the resulting equations which precludes their practical applications.
At least partly, this is because formulations of
Refs.\ \cite{kt08,andersson12,ach17,adhc17,ahdc17}  do not make a full use of
simplifications arising for a system,
for which the hydrodynamic approximation is valid, i.e., when the typical mean-free path $l$ and collision time $\tau$
are much smaller than, respectively, the typical length scale $L$ and timescale $\mathcal{T}$ of the problem.
Because the hydrodynamic approximation should work very well for typical NS conditions, it seems interesting and useful to formulate MHD
where this approximation is fully implemented.

Thus, the aim of the present study is to
 present a ready-to-use formulation of a dissipative relativistic MHD for multicomponent nonsuperfluid mixtures. To this aim,
 we also express the phenomenological coefficients
appearing in this hydrodynamics through
the parameters calculated from the 
microscopic theory.
We follow the textbook phenomenological approach of Landau and Lifshitz \cite{ll87} and Eckart \cite{eckart40},
namely, we build a first-order dissipative hydrodynamics which includes only the linear terms in the thermodynamic fluxes.
Notice that, theoretically speaking, the standard first-order hydrodynamics
has some issues with causality (thermal fluctuations propagate at infinite speed)
and stability \cite{hl83, hl85}.
These issues can be avoided at the cost of using much more complicated hydrodynamic theories,
such as various second-order theories \cite{israel76,is79,lsmr86}
or hydrodynamics obtained within Carter's variational approach \cite{carter91,andersson12,adhc17}
(see, e.g., Refs.\ \cite{ac07,rz13} for the reviews).%
\footnote{
	As has been shown recently, some first-order theories
	(which are more general than Landau-Lifshitz or Eckart hydrodynamics and contain additional kinetic coefficients)
	can also be stable and causal, at least in some cases~\cite{Van2012,kovtun19,bdn19}.
}\ %
However, the first-order theory is sufficient (and, in fact, follows from the kinetic equation approach: see, e.g., Ref.~\cite{DeGroot:1980dk})
	as long as the hydrodynamic description is applicable, that is,	$\mathcal{T}\gg\tau$  and $L \gg l$.
	These conditions ensure that the space-time gradients of the deviations from the equilibrium, e.g.,
	heat flux and viscous stress, are negligible on the scale of 
	mean-free path or mean-free time.
	Bearing in mind that these conditions hold for almost all situations of practical interest,
	such as magnetothermal evolution or hydrodynamic oscillations of NSs,
	we restrict ourselves to the first-order hydrodynamics.%
%
%%%%%%%
\footnote{Note that the instabilities of the first-order hydrodynamics found, e.g., in Ref.\ \cite{hl85},
		develop for the modes (Fourier components), which evolve on a timescale $\mathcal{T}$ much shorter
		than $\tau$. Such modes are not within the range of applicability of the first-order theory
		and should be discarded (e.g., \cite{shaefer14,kovtun19}) or filtered out,
		if we talk about numerical implementation of this theory.}
Another attractive feature of this hydrodynamics is that
it allows one to easily connect 
all the phenomenological kinetic coefficients with
the quantities calculated from the microscopic theory.
Finally,
the dissipative hydrodynamics presented below, 
combined with the results of Ref.~\cite{gd16},
is readily extendable
to the superfluid and superconducting charged mixtures with vortices \cite{dg20}.

This paper is organized as follows.
In Sec.~\ref{sec:gen-eqns},
we formulate general hydrodynamics equations
for charged relativistic mixtures
in the absence of bound charges and bound currents.
In Sec.~\ref{sec:entropy}, we derive the entropy generation
equation, and, in Sec.~\ref{sec:general},
we use it together with the Onsager relations
to derive the most general form of dissipative corrections
for particle currents and energy-momentum tensor.
In Sec.~\ref{sec:gr}, we explicitly write out the MHD equations
for a general relativistic nonrotating NS.
Section~\ref{sec:special-cases} contains a number of applications
and special cases that highlight the connection between 
the generalized diffusion coefficients, introduced here,
and kinetic coefficients known from the literature:
electrical conductivity, thermal conductivity,
and momentum transfer rates, as well as nonrelativistic diffusion, thermodiffusion and thermal conductivity coefficients.
Sec.~\ref{sec:Jik} contains the ready-to-use
expressions for the momentum transfer rates in $npe\mu$ matter of NS cores.
We compare our results to other works in Sec.~\ref{sec:toteq}, and sum up in Sec.~\ref{sec:conclusion}.
Some technical details elucidating derivations in the main text are presented
in Appendixes~\ref{sec:reaction-rates} and \ref{sec:gen-kinetic}.
Finally, Appendix~\ref{sec:ncomp} shows how to express generalized diffusion coefficients
through the momentum transfer rates in the low-temperature limit.

Unless otherwise stated, in what follows 
the speed of light $c$
and the Boltzmann constant $k_{\rm B}$ are set to unity:
$c=k_{\rm B}=1$.

%%%%%%%%%%%%%%%%%%%%%%%%%%%%%%%%%%%%%%%%%%%%%%%%%%%%%%%%%%%%%%%%%%%%
\section{General equations}
\label{sec:gen-eqns}
%%%%%%%%%%%%%%%%%%%%%%%%%%%%%%%%%%%%%%%%%%%%%%%%%%%%%%%%%%%%%%%%%%%%

In this section we present hydrodynamic equations
that describe 
charged relativistic mixtures
in the absence of bound charges and bound currents.
For the sake of simplicity here we assume that the space-time metric is flat:
$g_{\mu\nu}={\rm diag}(-1,1,1,1)$;
the straightforward generalization of our results to arbitrary $g_{\mu\nu}$ is discussed in Sec.~\ref{sec:gr}.
The hydrodynamic equations
include the energy-momentum conservation law
\begin{gather}
\label{eq:dTmunu=0}
	\partial_{\mu} T^{\mu\nu} = G^\nu
\end{gather}
and continuity equations for particle species $j$
\begin{equation}
\label{eq:jmu}
	\partial_\mu j_{(j)}^{\mu} = \Delta\Gamma_j
,
\end{equation}
where $\partial_\mu \equiv \partial/\partial x^{\mu}$;
$T^{\mu \nu}$ is the energy-momentum tensor 
(which must be symmetric)
$G^\nu$ is the radiation four-force density\footnote{
	$G^\nu$ describes the exchange of energy and momentum between matter and radiation.
	In the simplest case of isotropic emission, $G^\nu = - Q u^\nu$, where $Q$
	is the total emissivity
	(e.g., neutrino emissivity due to beta-processes in the NS cores).};
$j_{(j)}^{\mu}$ is the particle four-current density for the particle species $j$;
$\Delta\Gamma_j$ is the reaction rate for
species $j$ due to nonequilibrium processes of particle mutual transformation.
Here and below, unless otherwise stated, 
$\mu$, $\nu$, and other Greek letters
are space-time indices 
running over $0$, $1$, $2$, and $3$;
Latin letters $i,j,k\ldots$ are particle species indices,
and summation over repeated space-time and particle indices is assumed. 
Generally, $T^{\mu\nu}$ and $j_{(j)}^{\mu}$
can be presented as
\begin{eqnarray}
\label{eq:Tmunu}
	T^{\mu\nu} &=& {(P+\varepsilon) u^{\mu} u^\nu} + {P g^{\mu\nu}} + \Delta T^{\mu\nu}_{({\rm EM})} + \Delta \tau^{\mu\nu}
,\\
\label{eq:jmu2}
	j_{(j)}^{\mu} &=& {n_j u^{\mu}} + \Delta j_{(j)}^{\mu},
\end{eqnarray}
where $P$ is the pressure given by Eq.~\eqref{eq:pres} below; 
$\varepsilon$ is the energy density; 
$n_j$ is the number density for species $j$; 
$\Delta T^{\mu\nu}_{({\rm EM})}$
is the electromagnetic contribution to the energy-momentum tensor given by Eq.~\eqref{eq:dT-EM} below;
$\Delta \tau^{\mu\nu}$ and $\Delta j_{(j)}^{\mu}$
are dissipative corrections to the energy-momentum tensor and particle currents, respectively.
%$g_{\mu\nu}={\rm diag}(-1,1,1,1)$ is the space-time metric.\footnote{Straightforward generalization of our results to arbitrary $g_{\mu\nu}$ is discussed in Section~\ref{sec:gr}.}
%
Finally, $u^{\mu}$ is the four-velocity vector,
normalized by the condition
\begin{gather}
\label{eq:norm1}
	u_{\mu} u^{\mu} = -1.
\end{gather}
The thermodynamic quantities introduced in Eqs.~\eqref{eq:Tmunu} and \eqref{eq:jmu2}
do not have any direct physical meaning 
unless a frame 
where they are measured (defined) is specified.
In what follows we measure all thermodynamic quantities
in the comoving frame given by the condition 
$u^{\mu}=(1,0,0,0)$.
By definition, the energy and number densities are the components $T^{00}$ and $j_{(j)}^{0}$ in that frame, 
 $T^{00}=\varepsilon$ and $j^{0}_{(j)}=n_j$,
which, in an arbitrary frame, translate into
\begin{eqnarray}
\label{eq:condTmunu}
	u_{\mu}u_{\nu} T^{\mu\nu} &=& \varepsilon,
\\
\label{eq:condjmu}
	u_{\mu} j_{(j)}^{\mu} &=& -n_j.
\end{eqnarray}
These relations
in view of the expressions \eqref{eq:Tmunu}, \eqref{eq:jmu2}, and \eqref{eq:dT-EM}--\eqref{eq:TM}, impose the following restrictions on the dissipative corrections:
\begin{eqnarray}
\label{eq:condTmunu2}
	u_\mu u_{\nu} \Delta \tau^{\mu\nu} &=& 0
,\\
\label{eq:condjmu2}
	u_{\mu} \Delta j_{(j)}^{\mu} &=& 0.
\end{eqnarray}
The definition of $u^\mu$ is still somewhat ambiguous.
Following Landau and Lifshitz \cite{ll87},
we specify $u^\mu$ by the condition that 
the total momentum of the fluid vanishes in the comoving frame
(the so-called Landau-Lifshitz, or transverse, frame).
Then the dissipative correction to the energy-momentum in this frame obeys the additional restriction
\begin{gather}
\label{eq:LL-dt}
	u_{\nu} \Delta \tau^{\mu\nu} = 0
.
\end{gather}

In the system without bound charges and currents the electromagnetic field 
is described by Maxwell's equations
\begin{gather}
\label{eq:maxwell-1}
	\partial_{\mu} F_{\nu\lambda}
	+ \partial_{\nu} F_{\lambda\mu}
	+ \partial_{\lambda} F_{\mu\nu}
	= 0
,\\
\label{eq:maxwell-2}
	\partial_\nu F^{\mu\nu}
	= 4\pi J^{\mu}
,
\end{gather}
where $F_{\mu\nu} = - F_{\nu\mu}$ is the electromagnetic tensor,
and
\begin{gather}
\label{eq:Jfree}
	J^{\mu}
	\equiv e_j j_{(j)}^\mu
	= e_j n_j u^\mu +e_j \Delta j_{(j)}^\mu
\end{gather}
is the electric current ($e_j$ is the electric charge for particle species $j$).
Since $\Delta j_{(j)}^\mu$ 
depends on the electric four-vector $E^{\mu}$ and
various gradients of chemical potentials and temperature
[see Eq.\ (\ref{eq:djmu2}) below],
%Consequently,
Eq.\ (\ref{eq:Jfree}) is simply the generalized Ohm's law,
whose form is specified in Sec.\ \ref{sec:general}.
Note that in the majority of situations the term $e_j n_j u^{\mu}$
in Eq.\ (\ref{eq:Jfree}) is very small in comparison to $e_j \Delta j_{(j)}^\mu$ due to quasineutrality
condition $e_j n_j\approx 0$ and can be neglected \cite{brag65}.
The electric field ${\pmb E}$ and the magnetic induction ${\pmb B}$ 
are defined, respectively, as
\begin{gather}
\label{eq:E-3d}
    {\pmb E} \equiv \left( F^{01}, F^{02}, F^{03} \right)
,\\
\label{eq:B-3d}
    {\pmb B} \equiv \left( F^{23}, F^{31}, F^{12} \right)
.
\end{gather}

Eqs.~\eqref{eq:dTmunu=0}--\eqref{eq:Tmunu}, \eqref{eq:condTmunu}, and \eqref{eq:condjmu}
are key equations that will be used below.
They should be supplemented by the second law of thermodynamics,
\begin{equation}
\label{eq:2ndlaw}
	d \varepsilon = \mu_j \, dn_j 
		+ T \, dS 
		+ d \varepsilon_{\rm add}
,
\end{equation}
where $\mu_j$ is the relativistic chemical potential
for particle species $j$,
$T$ is the temperature,
$S$ is the entropy per unit volume, and $d \varepsilon_{\rm add}$ is the electromagnetic contribution.
In the absence of bound charges and currents,
%when ${\pmb E} = {\pmb D}$ and ${\pmb B} = {\pmb H}$,
the latter reads
\begin{gather}
\label{eq:de-EM}
	d \varepsilon_{\rm add}
	= \frac{1}{4\pi} E_\alpha dE^\alpha + \frac{1}{4\pi} B_\alpha dB^\alpha 
,
\end{gather}
where $E^\alpha$ and $B^\alpha$ are the ``electric'' and ``magnetic''
four-vectors, respectively \cite{lichnerowicz67,gd16},
defined as 
\begin{gather}
\label{eq:Emu}
	E^\mu \equiv u_\nu F^{\mu\nu}
,\\
\label{eq:Bmu}
	B^\mu \equiv \frac{1}{2} \epsilon^{\mu\nu\alpha\beta} u_\nu F_{\alpha\beta}
,
\end{gather}
and $\epsilon^{\mu\nu\alpha\beta}$ is the Levi-Civita tensor, normalized by $\epsilon^{0123}=1$.
In the comoving frame, $u^\mu = (1,0,0,0)$,
four-vectors $E^\mu$ and $B^\mu$ are expressed through
the electric field ${\pmb E}$
and the magnetic induction ${\pmb B}$,
respectively:
$E^\mu = (0, {\pmb E})$ and
$B^\mu = (0, {\pmb B})$.

The pressure $P$
is defined as a partial derivative of the full system energy
$\varepsilon V$ with respect to volume $V$ at a constant 
total number of particles, total entropy, and electromagnetic scalars $E_\alpha E^\alpha$ and $B_\alpha B^\alpha$:
\begin{gather}
\label{eq:pres}
	P \equiv -\frac{\partial \left(\varepsilon V \right)}{\partial V}
	= -\varepsilon +\mu_j n_j + TS.
\end{gather}
Using Eqs.~\eqref{eq:2ndlaw}--\eqref{eq:pres},
one arrives at the following Gibbs-Duhem equation:
\begin{equation}
\label{eq:dP}
	dP = n_j  \, d\mu_j + S \, dT
	- \frac{1}{4\pi} E_\alpha dE^\alpha 
	- \frac{1}{4\pi} B_\alpha dB^\alpha
.
\end{equation}
The term 
$\Delta T^{\mu\nu}_{({\rm EM})}$ in Eq.~\eqref{eq:Tmunu}
has the form
\begin{gather}
\label{eq:dT-EM}
	\Delta T^{\mu\nu}_{({\rm EM})}
	= T^{\mu\nu}_{({\rm E})}
	+ T^{\mu\nu}_{({\rm M})}
,
\end{gather}
where
$T^{\mu\nu}_{({\rm E})}$ and $T^{\mu\nu}_{({\rm M})}$
are given, respectively, by [see, e.g., Eqs. (48) and (49) in Ref.\ \cite{gd16}]
\begin{gather}
\label{eq:TE}
	T^{\mu\nu}_{({\rm E})}
	= -\frac{1}{4\pi} \left(E^\mu E^\nu - \perp^{\mu\nu} E_\alpha E^\alpha \right)
,\\
\label{eq:TM}
	T^{\mu\nu}_{({\rm M})}
	= \frac{1}{4\pi} \left(
			\perp_{\delta \alpha} F^{\mu\delta} F^{\nu\alpha}
			- u^\mu u^\nu u^\gamma u_\beta F^{\alpha\beta} F_{\alpha\gamma}
		\right)
,
\end{gather}
where $\perp^{\mu\nu} \equiv g^{\mu\nu} + u^\mu u^\nu$ is the projection tensor.

%%%%%%%%%%%%%%%%%%%%%%%%%%%%%%%%%%%%%%%%%%%%%%%%%%%%%%%%%%%%%%%%%%%%%%%%%%%%
\section{Entropy generation rate}
\label{sec:entropy}
%%%%%%%%%%%%%%%%%%%%%%%%%%%%%%%%%%%%%%%%%%%%%%%%%%%%%%%%%%%%%%%%%%%%%%%%%%%%

The entropy generation equation follows naturally from the conservation laws and the second law of thermodynamics. Consider the combination $u_\nu \partial_{\mu} T^{\mu\nu} - u_\nu G^\nu$,
which vanishes in view of Eq.~\eqref{eq:dTmunu=0}.
Using Eqs.~\eqref{eq:jmu}--\eqref{eq:jmu2},
\eqref{eq:2ndlaw}, and \eqref{eq:pres},
as well as the identities
$u_\nu \partial_\mu u^\nu = 0$ and $\partial_{\mu} g^{\mu\nu} = 0$,
we arrive at
\begin{gather}
\label{eq:dSmu1}
	\partial_{\mu} \left(S u^\mu \right)
	= \frac{\mu_j}{T} \partial_{\mu} \Delta j_{(j)}^\mu
	- \frac{\mu_j}{T} \Delta\Gamma_j
	- \frac{u^\mu}{T} \partial_{\mu} \varepsilon_{\rm add}
	+ \frac{u_\nu}{T} \partial_{\mu} \left( \Delta T^{\mu\nu}_{({\rm EM})} + \Delta \tau^{\mu\nu}\right)
	- \frac{Q}{T}
,
\end{gather}
where $Q \equiv u_\nu G^\nu$.
Now let us transform 
the ``electromagnetic'' term $u^\mu \partial_{\mu}\varepsilon_{\rm add}/T$ in Eq.\ \eqref{eq:dSmu1}.
Using Eqs.~(63) and (68) from Ref.~\cite{gd16}, we can
write
\begin{gather}
\label{eq:udeEM}
	-u^{\mu} \partial_{\mu}\varepsilon_{\rm add}
	= u^{\nu} F_{\mu\nu} \, J^{\mu}
	- \partial_\mu \left[
			u^\nu \left( T^\mu_{({\rm E}) \, \nu}
			+ T^\mu_{({\rm M}) \, \nu} 
		\right)\right]
	+ \partial_{\mu} u^{\nu}\left( 
	T^\mu_{({\rm E}) \, \nu}+
	T^\mu_{({\rm M}) \, \nu} 
	\right)
.
\end{gather}
Notice that
$u^{\nu} F_{\mu\nu} J^{\mu}
	= e_j E_\mu \Delta j_{(j)}^\mu$
due to antisymmetry of the tensor $F^{\mu\nu}$
[remember also the definition \eqref{eq:Emu} for $E^\mu$].
Substituting now relations~\eqref{eq:dT-EM} and \eqref{eq:udeEM}
into Eq.~\eqref{eq:dSmu1}, we obtain
\begin{gather}
\label{eq:dSmu-EM}
	\partial_{\mu} \left(S u^\mu \right)
	= \frac{\mu_j}{T} \partial_{\mu} \Delta j_{(j)}^\mu
	- \frac{\mu_j}{T} \Delta\Gamma_j
	+ \frac{e_j E_\mu}{T}  \Delta j_{(j)}^\mu
	+ \frac{u_\nu}{T} \partial_{\mu} \Delta \tau^{\mu\nu}
	- \frac{Q}{T}
%,
\end{gather}
or, equivalently,
\begin{gather}
\label{eq:dSmu-EM2}
	\partial_{\mu} \left(
		S u^\mu
		- \frac{\mu_j}{T} \Delta j_{(j)}^\mu
		- \frac{u_\nu}{T} \Delta \tau^{\mu\nu}
		\right)
	= - \Delta j_{(j)}^\mu \left[
				\partial_{\mu} \left(\frac{\mu_j}{T}\right) 
				- \frac{e_j E_\mu}{T}
			\right]
		- \Delta \tau^{\mu\nu} \partial_{\mu} \left(\frac{u_\nu}{T}\right) 
		- \frac{\mu_j}{T} \Delta\Gamma_j
		- \frac{Q}{T}
.
\end{gather}
The left-hand side of this equation is the four-divergence
of the entropy four-current
$S^\mu = S u^\mu
		- \frac{\mu_j}{T} \Delta j_{(j)}^\mu
		- \frac{u_\nu}{T} \Delta \tau^{\mu\nu}$,
while the right-hand side represents the entropy generation
due to dissipative processes.
Note also that dissipative corrections to the currents $\Delta j_{(j)}^\mu$
and the energy-momentum tensor $\Delta \tau^{\mu\nu}$
must be expressed as linear combinations of gradients
of $u^\mu$ and thermodynamic variables \cite{weinberg71}.\footnote{
We remind the reader that we restrict ourselves to the first-order
dissipative hydrodynamics.
}
Except for the last term, which can be arbitrary, the right-hand side of Eq.~\eqref{eq:dSmu-EM2} must be non-negative
for all possible fluid configurations.

In the Landau-Lifshitz frame defined by Eq.~\eqref{eq:LL-dt},
Eq.~\eqref{eq:dSmu-EM2} reduces to
\begin{gather}
\label{eq:dSmu-EM-LL-0}
	\partial_{\mu} S^\mu
	= - \Delta j_{(j)}^\mu \left[
				\partial_{\mu} \left(\frac{\mu_j}{T}\right) 
				- \frac{e_j E_\mu}{T}
			\right]
		- \Delta \tau^{\mu\nu} \frac{\partial_{\mu} u_\nu}{T} 
		- \frac{\mu_j}{T} \Delta\Gamma_j
		- \frac{Q}{T}
,\quad
	S^\mu = S u^\mu
		- \frac{\mu_j}{T} \Delta j_{(j)}^\mu
.
\end{gather}
Using the conditions \eqref{eq:condjmu2} and \eqref{eq:LL-dt},
one can rewrite Eq.~\eqref{eq:dSmu-EM-LL-0} as
\begin{gather}
\label{eq:dSmu-EM-LL}
	\partial_{\mu} S^\mu
	= - \Delta j_{(j)}^\mu d_{(j)\mu}
		- \Delta \tau^{\mu\nu} \frac{\nablaperp_{\mu} u_\nu}{T} 
		- \frac{\mu_j}{T} \Delta\Gamma_j
		- \frac{Q}{T}
,
\end{gather}
where we introduced the orthogonal
part of the four-gradient
\begin{gather}
\label{eq:nabla}
	\nablaperp_{\mu} \equiv \perp_{\mu\nu} \partial^\nu	
,
\end{gather}
and the vector $d_{(j)\mu}$,
\begin{gather}
\label{eq:dmu}
	d_{(j)\mu} \equiv
	\nablaperp_{\mu} \left(\frac{\mu_j}{T}\right) 
					- \frac{e_j E_\mu}{T}
.
\end{gather}
Both the vector $d_{(j)\mu}$ and the tensor $\nablaperp_{\mu} u_\nu$
are orthogonal to the four-velocity $u^\mu$.

The term $-\frac{\mu_j}{T} \Delta\Gamma_j$ in Eq.~\eqref{eq:dSmu-EM-LL} can be rewritten in the form (see Appendix~\ref{sec:reaction-rates})
\begin{gather}
\label{eq:muDeltaGamma}
	-\frac{\mu_j}{T} \Delta\Gamma_j
	= \frac{1}{T} \lambda_X \left( \Delta\mu_X \right)^2
,
\end{gather}
where $\Delta\mu_X$
is the chemical potential imbalance
for a given reaction $X$ (for example, for the direct or modified Urca processes \cite{ykgh01},
the chemical potential imbalance equals $\Delta\mu_X \equiv \mu_{n} - \mu_{p} - \mu_{e}$)
and $\lambda_X > 0$ is a corresponding reaction coefficient.

%%%%%%%%%%%%%%%%%%%%%%%%%%%%%%%%%%%%%%%%%%%%%%%%%%%%%%%%%%%%%%%%%%%%%%%%%%%%%%%%%
\section{Expressions for $\Delta j_{(j)}^\mu$ and $\Delta \tau^{\mu\nu}$}
\label{sec:general}
%%%%%%%%%%%%%%%%%%%%%%%%%%%%%%%%%%%%%%%%%%%%%%%%%%%%%%%%%%%%%%%%%%%%%%%%%%%%%%%%%

In the linear approximation in small gradients,
the quantities
$\Delta j_{(j)}^\mu$ and $\Delta T^{\mu\nu}$ can generally be presented as 
\begin{gather}
\label{eq:dj}
	\Delta j_{(j)}^\mu
		= - A_{jk}^{\mu\nu} d_{(k)\nu}
		  - B_j^{\mu\nu\lambda} \frac{\nablaperp_\nu u_\lambda}{T}
,\\
\label{eq:dtau}
	\Delta \tau^{\mu\nu}
		= - C_k^{\mu\nu \lambda} d_{(k)\lambda}
		  - D^{\mu\nu \lambda \sigma} \nablaperp_\lambda u_\sigma
,
\end{gather}
respectively,
where the kinetic coefficients $A_{jk}^{\mu\nu}$, $B_j^{\mu\nu\lambda}$, $C_k^{\mu\nu \lambda}$, and $D^{\mu\nu \lambda \sigma}$
are discussed in 
%the next section.
what follows.

%%%%%%%%%%%%%%%%%%%%%%%%%%%%%%%%%%%%%%%%%%%%%%%%%%%%%%%%%%%%%%%%%%%%%%%%%%%%%%%%%
\subsection{No magnetic field}
\label{sec:general:no-B}
%%%%%%%%%%%%%%%%%%%%%%%%%%%%%%%%%%%%%%%%%%%%%%%%%%%%%%%%%%%%%%%%%%%%%%%%%%%%%%%%%

Let us first analyze expressions
for $\Delta j_{(j)}^\mu$ and $\Delta \tau^{\mu\nu}$
in a homogeneous matter in the absence of a preferred direction (e.g., the magnetic field).
In this case the kinetic coefficients $A_{jk}^{\mu\nu}$, $B_j^{\mu\nu\lambda}$, $C_k^{\mu\nu \lambda}$, and $D^{\mu\nu \lambda \sigma}$
in Eqs.~\eqref{eq:dj} and \eqref{eq:dtau}
depend only on $u^\mu$, $g^{\mu\nu}$, and (scalar) equilibrium thermodynamic functions.
These coefficients meet a number of conditions.
First, application of the Onsager symmetry principle gives
\begin{gather}
\label{eq:Ajk=Akj}
	A_{jk}^{\mu\nu} = A_{kj}^{\nu\mu}
,\\
\label{eq:Dmnls=Dlsmn}
	D^{\mu\nu\lambda\sigma}  = D^{\lambda\sigma\mu\nu}
,\\
\label{eq:C=B}
	C_k^{\mu\nu \lambda} = - B_k^{\lambda\mu\nu}
.
\end{gather}
Second, the symmetry of $\Delta \tau^{\mu\nu}$ implies that
\begin{gather}
\label{eq:Dmnls=Dnmls}
	D^{\mu\nu\lambda\sigma}  = D^{\nu\mu\lambda\sigma}
,\\
\label{eq:Cmnl=Cnml}
	C_k^{\mu\nu\lambda} = C_k^{\nu\mu\lambda}
.
\end{gather}
Third, the conditions \eqref{eq:condjmu2} and \eqref{eq:LL-dt}
lead to the following constraints:
\begin{gather}
\label{eq:uA=0}
	u_\mu A_{jk}^{\mu\nu} d_{(k)\nu} = 0
,\\
\label{eq:uB=0}
	u_\mu B_{k}^{\mu\nu\lambda} \nablaperp_\nu u_\lambda = 0
,\\
\label{eq:uC=0}
	u_\mu C_{k}^{\mu\nu\lambda} d_{(k)\lambda} = 0
,\\
\label{eq:uD=0}
	u_\mu D^{\mu\nu\lambda\sigma} \nablaperp_\lambda u_\sigma = 0
,
\end{gather}
which must hold for arbitrary $d_{(k)\nu}$ and $\nablaperp_\nu u_\lambda$.
Finally, one should require that the matrices of kinetic coefficients
should be such that 
the entropy production rate
would be non-negative.

Unless the isotropic system has no center of inversion, the Curie principle requires that the perturbations of different tensor structure (i.e. viscosity and diffusion) do not interfere and $B_{k}^{\mu\nu\lambda} = C_{k}^{\mu\nu\lambda} = 0$ (see also Appendix~\ref{sec:gen-kinetic}). 
Then the most general expressions for $\Delta j_{(j)}^\mu$ and $\Delta \tau^{\mu\nu}$ take the form
[see Eqs.~(\ref{eq:app:Ajk-noB}) and (\ref{eq:app:Dmnls-noB})]
\begin{gather}
\label{eq:dj-noB}
	\Delta j_{(j)}^\mu
		= - \mathcal{D}_{jk} d_{(k)}^\mu 
,\\
	\Delta \tau^{\mu\nu}
	= - \eta \left(
			   \nablaperp^\mu u^\nu
			+ \nablaperp^\nu u^\mu
		\right)
	- \left(\zeta - \frac{2}{3} \eta \right)
		\perp^{\mu\nu}
		\nablaperp_\lambda u^\lambda
,
\end{gather}
respectively.
Here the matrix of generalized diffusion coefficients $\mathcal{D}_{jk}$ must be positive definite,
and the coefficients $\eta$ (shear viscosity) and $\zeta$ (bulk viscosity) \cite{ll87} must be non-negative.

%%%%%%%%%%%%%%%%%%%%%%%%%%%%%%%%%%%%%%%%%%%%%%%%%%%%%%%
\subsection{Accounting for the magnetic field}
\label{sec:general:B}
%%%%%%%%%%%%%%%%%%%%%%%%%%%%%%%%%%%%%%%%%%%%%%%%%%%%%%%

Now let us consider a homogeneous matter,
in which the only preferred direction is specified, in the comoving frame, 
by the magnetic induction vector ${\pmb B}$.%
\footnote{As in the ordinary MHD (see, e.g., Ref.~\cite{ll60}, Sec.~66 and Ref.~\cite{ll80}, Sec.~58),	
	the electric field in that frame is assumed to be sufficiently small, of the order of gradients of thermodynamic functions, and does not provide an additional preferred direction in the system.}
In what follows, in addition to the magnetic four-vector $B^\mu$ it will be convenient to use also
the tensor
$\Fperp^{\mu\nu} \equiv \perp^{\mu\alpha} \perp^{\nu\beta} F_{\alpha\beta}$ (see Ref.\ \cite{gd16}, Appendix A),
which in the comoving frame is given by
\begin{gather}
\Fperp^{\mu\nu}
=
\left(
\begin{array}{cccc}
	0 & 0 & 0 & 0
	\\
	0 & 0 & B_3 & -B_2
	\\
	0 & -B_3 & 0 & B_1
	\\
	0 & B_2 & -B_1 & 0
\end{array}
\right)
\end{gather}
and satisfies the identity
\begin{gather}
\label{eq:uFperp=0}
	u_\nu \Fperp^{\mu\nu} = 0
.
\end{gather}

In the presence of a magnetic field, the Onsager principle \eqref{eq:Ajk=Akj}--\eqref{eq:C=B} is modified
(see, e.g., \cite{ll5}, Sec.~120):
\begin{gather}
\label{eq:B:Ajk=Akj}
	A_{jk}^{\mu\nu}({\pmb B}) = A_{kj}^{\nu\mu} (-{\pmb B})
,\\
\label{eq:B:Dmnls=Dlsmn}
	D^{\mu\nu\lambda\sigma}({\pmb B})  = D^{\lambda\sigma\mu\nu}(-{\pmb B})
,\\
\label{eq:B:C=B}
	C_k^{\mu\nu \lambda}({\pmb B}) = - B_k^{\lambda\mu\nu}(-{\pmb B})
,
\end{gather}
while
Eqs.~\eqref{eq:dj} and \eqref{eq:dtau} and
the conditions \eqref{eq:Dmnls=Dnmls}--\eqref{eq:uD=0} remain unaffected.
The kinetic coefficients now depend on
$u^\mu$, $g^{\mu\nu}$, $B^\mu$, $\Fperp^{\mu\nu}$, and (scalar) equilibrium thermodynamic functions.

 One can check (see Appendix~\ref{sec:gen-kinetic}) that, as in the system without the magnetic field, $B_{k}^{\mu\nu\lambda} = C_{k}^{\mu\nu\lambda} = 0$;
i.e., 
diffusion and viscosity do not interfere.
Thus, $\Delta j_{(j)}^\mu$ depends only on the vectors $d_{(k)\nu}$.
In the comoving frame
$d_{(k)}^\mu = (0, {\pmb d}_{(k)})$,
$\Delta j_{(j)}^\mu = (0, {\Delta \pmb j}_{(j)})$ [see condition~\eqref{eq:condjmu2}];
thus, ${\Delta \pmb j}_{(j)}$ can be generally presented as (see Appendix B)
\begin{gather}
\label{eq:dj-comoving2}
	{\Delta \pmb j}_{(j)}
	= - \mathcal{D}_{jk}^{\parallel} {\pmb d}_{(k)\parallel}
	- \mathcal{D}_{jk}^{\perp} {\pmb d}_{(k)\perp}
	- \mathcal{D}_{jk}^{H} \left[ {\pmb d}_{(k)\perp} \times {\pmb b} \right]
.
\end{gather}
Here
$\mathcal{D}_{jk}^{\parallel}$, $\mathcal{D}_{jk}^{\perp}$,
and $\mathcal{D}_{jk}^{H}$ are the diffusion coefficients;
${\pmb b} \equiv {\pmb B}/ |{\pmb B}|$
is the unit vector in the direction of the magnetic field;
and the vectors ${\pmb d}_{(k)\parallel}$ and ${\pmb d}_{(k)\perp}$
are defined, respectively, as
\begin{gather}
	{\pmb d}_{(k)\parallel} \equiv ({\pmb d}_{(k)} {\pmb b}) {\pmb b}
,\\
	{\pmb d}_{(k)\perp}
	\equiv {\pmb d}_{(k)} - ({\pmb d}_{(k)} {\pmb b}) {\pmb b}
	= {\pmb b} \times \left[ {\pmb d}_{(k)} \times {\pmb b} \right]
.
\end{gather}
In an arbitrary frame, Eq.~\eqref{eq:dj-comoving2} can be rewritten as
\begin{gather}
\label{eq:djmu2}
	\Delta j_{(j)}^\mu
	=
	- \mathcal{D}_{jk}^{\parallel} b^\mu b^\nu d_{(k)\nu}
	- \mathcal{D}_{jk}^{\perp} \left( \perp^{\mu\nu} - b^\mu b^\nu \right) d_{(k)\nu}
	- \mathcal{D}_{jk}^{H} b^{\mu\nu} d_{(k)\nu}
,
\end{gather}
where we introduced\footnote{
Note that $d_{(k)\nu} \Fperp^{\mu\nu} = \left(0, \left[ {\pmb d}_{(k)} \times {\pmb B} \right] \right)$
in the comoving frame.
}
\begin{gather}
	b^\mu \equiv \frac{B^\mu}{\sqrt{B_\alpha B^\alpha}}
,\\
	b^{\mu\nu} \equiv \frac{\Fperp^{\mu\nu}}{\sqrt{B_\alpha B^\alpha}}
.
\end{gather}

In the absence of viscosity, chemical reactions, and energy losses ($Q=0$),
the entropy generation equation \eqref{eq:dSmu-EM-LL} reduces to
\begin{gather}
	\partial_{\mu} S^\mu
	= \mathcal{D}_{jk}^{\parallel} {\pmb d}_{(j)\parallel} {\pmb d}_{(k)\parallel}
	+ \mathcal{D}_{jk}^{\perp} {\pmb d}_{(j)\perp} {\pmb d}_{(k)\perp}
 	= \mathcal{D}_{jk}^{\parallel} d_{(j)\parallel \mu} d_{(k)\parallel}^\mu
	+ \mathcal{D}_{jk}^{\perp} d_{(j)\perp \mu} d_{(k)\perp}^\mu
,
\end{gather}
where
\begin{gather}
	d_{(k)\parallel}^\mu
		\equiv	b^\mu b^\nu d_{(k)\nu}
,\quad
	d_{(k)\perp}^\mu
	\equiv	\left( \perp^{\mu\nu} - b^\mu b^\nu \right) d_{(k)\nu}
.
\end{gather}
As follows from the Onsager principle~\eqref{eq:B:Ajk=Akj},
matrices $\mathcal{D}_{jk}^{\parallel}$, $\mathcal{D}_{jk}^{\perp}$,
and $\mathcal{D}_{jk}^{H}$ must be symmetric.
In addition, $\mathcal{D}_{jk}^{\parallel}$ and $\mathcal{D}_{jk}^{\perp}$
must also be positive definite in order to ensure that the entropy of the system does not decrease.
In the limit ${\pmb B} \to 0$, one has
$\mathcal{D}_{jk}^{\parallel} = \mathcal{D}_{jk}^{\perp} = \mathcal{D}_{jk}$,
and $\mathcal{D}_{jk}^{H} = 0$
[see Eq.~(\ref{eq:dj-noB})].

Since the coefficient $C_j^{\mu\nu\lambda}$ vanishes,
the general expression for $\Delta\tau^{\mu\nu}$
reads [see Eq.~(\ref{eq:app:Dmlns-final})]
\begin{gather}
\label{eq:dtmn2}
\begin{split}
	\Delta\tau^{\mu\nu}
	= &-\frac{1}{3}\eta_0 \left[
			\Xi^{\mu\nu} \Xi^{\lambda\sigma}
			- 2 \Xi^{\mu\nu} b^\lambda b^\sigma
			- 2 \Xi^{\lambda\sigma} b^\mu b^\nu
			+ 4 b^\mu b^\nu b^\lambda b^\sigma
			\right] \nablaperp_\lambda u_\sigma
	\\&- \eta_1 \left[
				\Xi^{\mu\lambda} \Xi^{\nu\sigma} 
				+ \Xi^{\mu\sigma} \Xi^{\nu\lambda}
				- \Xi^{\mu\nu} \Xi^{\lambda\sigma} 
			\right] \nablaperp_\lambda u_\sigma
	\\&- \eta_2 \left[
			\Xi^{\mu\lambda} b^\nu b^\sigma
			+ \Xi^{\mu\sigma} b^\nu b^\lambda
			+ \Xi^{\nu\lambda} b^\mu b^\sigma
			+ \Xi^{\nu\sigma} b^\mu b^\lambda
		\right] \nablaperp_\lambda u_\sigma
	\\&- \frac{1}{2} \eta_3 \left[
			\Xi^{\mu\lambda} b^{\nu\sigma}
			+ \Xi^{\mu\sigma} b^{\nu\lambda}
			+ \Xi^{\nu\lambda} b^{\mu\sigma}
			+ \Xi^{\nu\sigma} b^{\mu\lambda}
		\right] \nablaperp_\lambda u_\sigma
	\\&- \eta_4 \left[
			b^\mu b^\lambda b^{\nu\sigma}
			+ b^\mu b^\sigma b^{\nu\lambda}
			+ b^\nu b^\lambda b^{\mu\sigma}
			+ b^\nu b^\sigma b^{\mu\lambda}
		\right] \nablaperp_\lambda u_\sigma
	\\&- \zeta \perp^{\mu\nu} \perp^{\lambda\sigma}
			\nablaperp_\lambda u_\sigma
	\\&- \zeta_1 \left[
				\perp^{\mu\nu} b^\lambda b^\sigma
				+\perp^{\lambda\sigma} b^\mu b^\nu
			\right] \nablaperp_\lambda u_\sigma
,
\end{split}
\end{gather}
where $\Xi^{\mu\nu} \equiv \perp^{\mu\nu} - b^\mu b^\nu$. The quantities
$\eta_0 \ldots \eta_4$ are five shear viscosity coefficients,
and $\zeta$ and $\zeta_1$ are two bulk viscosity coefficients
\cite{lp87}. In the case of a vanishing external magnetic field, ${\pmb B}\to 0$, $\zeta_1 = 0$,
$\eta_0=\eta_1=\eta_2=\eta$, and $\eta_3=\eta_4=0$.
Equation~\eqref{eq:dtmn2} is a relativistic generalization
of the nonrelativistic expression for the stress tensor in the magnetic field,
which contains the same number of shear and bulk viscosity coefficients \cite{lp87}.
Phenomenological Eqs.~\eqref{eq:djmu2} and \eqref{eq:dtmn2} are also compatible with the results of the relativistic kinetic theory \cite{VanErkelens1977PhyA-1,vanErklens1977PhyA-2,vanErkelens1978}.

%%%%%%%%%%%%%%%%%%%%%%%%%%%%%%%%%%%%%%%%%%%%%%%%%%%%%%%%%%%%%%%%%%%%
\section{Accounting for general relativity}
\label{sec:gr}
%%%%%%%%%%%%%%%%%%%%%%%%%%%%%%%%%%%%%%%%%%%%%%%%%%%%%%%%%%%%%%%%%%%%

In the previous sections we assumed that the metric is flat,
$g_{\mu\nu}={\rm diag}(-1,1,1,1)$.
Generalization of our results to arbitrary $g_{\mu\nu}$ is straightforward
provided that all relevant length scales in the problem 
(e.g., particle mean-free paths)
are small enough compared with the 
characteristic gravitational length scale (e.g., NS radius) \cite{weinberg72}.
In the latter case the general relativity effects can be easily incorporated into hydrodynamics 
by replacing ordinary derivatives in all equations with their covariant analogs
and by replacing the Levi-Civita tensor $\epsilon^{\mu\nu\lambda\sigma}$ with 
$\eta^{\mu\nu\lambda\sigma} \equiv \sqrt{-{\rm det}~g}~ \epsilon^{\mu\nu\lambda\sigma}$.

In this section, we explicitly write out a set of general relativistic MHD equations
for a spherically symmetric star,
with the metric
\begin{gather}
	ds^2 = - e^\nu dt^2 + e^\lambda dr^2 + r^2 d\theta^2 + r^2 \sin^2 \theta d\phi^2
.
\end{gather}
We ignore metric perturbations caused by the magnetic field and fluid motions and work in the linear order in dissipative terms 
and velocities;
in particular, we neglect the terms like $|{\pmb u}|^2$, ${\pmb u} \Delta {\pmb j}_{(j)}$, and $[{\pmb u} \times {\pmb E}]$.
We also ignore effects of viscosity, which can be easily incorporated when needed.
For definiteness, we present the hydrodynamic equations for an NS core
consisting of neutrons ($n$), protons ($p$), electrons ($e$), and muons ($\mu$).
We take into account nonequilibrium direct and modified Urca processes \cite{ykgh01}, 
as well as energy losses due to isotropic neutrino emission with the emissivity $Q$.

In what follows, all three-vector components are measured by a static local observer,
in a locally flat frame (denoted by a hat)\footnote{
In other words,
${\pmb X}
\equiv \left( X^{\hat{r}}, X^{\hat{\theta}}, X^{\hat{\phi}} \right)
 = \left( {\rm e}^{\lambda/2} X^r, X^\theta, X^\phi \right)
$,
where ${\pmb X}$ is an arbitrary three-vector
and $\left( X^r, X^\theta, X^\phi \right)$ are its components
measured by a distant observer.
}
\begin{gather}
	d\hat{x}^\mu =  \left( d\hat{t}, d\hat{r}, d\hat{\theta}, d\hat{\phi} \right)
	= \left( {e}^{\nu/2} dt,  {\rm e}^{\lambda/2} dr, r d\theta, r \sin\theta d\phi \right)
.
\end{gather}
In other words, we introduce an orthonormal tetrad
%$e^a_{~\mu} = {\rm diag} \left( {e}^{-\nu/2},  {\rm e}^{-\lambda/2}, \frac{1}{r}, \frac{1}{r\sin\theta}\right)$
carried by the observer
and describe physical quantities by their projections on this tetrad (see, e.g., Refs.~\cite{tm82,bpt72}).

The three-vector
${\pmb u}$
is composed of spatial components of the four-velocity $u^\mu$,
${\pmb u} = \left( u^{\hat{r}}, u^{\hat{\theta}}, u^{\hat{\phi}} \right)$.
In the linear approximation, it
%is related to
can be expressed through
the ordinary three-dimensional velocity
${\pmb v} = \left( 
	\frac{{\rm d} r}{{\rm d} t},
	r \frac{{\rm d} \theta}{{\rm d} t},
	r \sin\theta \frac{{\rm d} \phi}{{\rm d} t}
\right)$,
measured by a distant observer,
as
\begin{gather}
	{\pmb u}
	= \left( 
		{\rm e}^{\lambda/2 -\nu/2}  v^r,
		{\rm e}^{-\nu/2} v^\theta,
		{\rm e}^{-\nu/2} v^\phi
	\right)
.
\end{gather}

In the following equations, we also introduce
time and space derivatives
taken by the local observer:
\begin{gather}
\label{eq:gr:dt-hat}
	\pd{}{\hat{t}}
	\equiv 	{\rm e}^{-\nu/2} \pd{}{t}
,\\
\label{eq:gr:nabla-hat}
	\hat{\pmb \nabla}
	\equiv \left(
		{\rm e}^{-\lambda/2} \pd{}{r},
		\frac{1}{r} \pd{}{\theta},
		\frac{1}{r\sin\theta} \pd{}{\phi}
	\right)
.
\end{gather}

Using the above definitions,
Maxwell equations \eqref{eq:maxwell-1} and \eqref{eq:maxwell-2}
can be rewritten
in terms of the electric field ${\pmb E}$ \eqref{eq:E-3d} and the magnetic induction ${\pmb B}$ \eqref{eq:B-3d}
as follows\footnote{Similar equations can be found, e.g., in Ref.~\cite{ram2001} [see Eqs.~(30)--(37) there with $\omega=0$].}:
\begin{gather}
\label{eq:gr:divB}
	\hat{\pmb \nabla} \cdot {\pmb B} = 0
,\\
\label{eq:gr:rotE}
	\pd{{\pmb B}}{\hat{t}}
	= - {\rm e}^{-\nu/2} ~
		\hat{\pmb \nabla} \times
		\left( {\pmb E} ~{\rm e}^{\nu/2} \right)
,\\
\label{eq:gr:divE}
	\hat{\pmb \nabla} \cdot {\pmb E}
	= 4\pi  e_j n_j
,\\
\label{eq:gr:rotB}
	{\rm e}^{-\nu/2} ~
		\hat{\pmb \nabla} \times
		\left( {\pmb B} ~{\rm e}^{\nu/2} \right)
	= 4 \pi {\pmb J} - \pd{\pmb E}{\hat{t}}
,
\end{gather}
where
${\pmb J}
	\equiv e_j {\pmb j}_{(j)}
	= e (n_{p} - n_{e} - n_{\mu}) {\pmb u}
	+ e (\Delta{\pmb j}_{p} - \Delta{\pmb j}_{e} - \Delta{\pmb j}_{\mu})$
is the electric current.

Continuity equation \eqref{eq:jmu} for particle species $j$
reads
\begin{gather}
	\pd{n_j}{\hat{t}}
	  + {\rm e}^{-\nu/2} ~
			\hat{\pmb \nabla}
			\left[
				\left( n_j {\pmb u} + \Delta {\pmb j}_{(j)} \right)
				{\rm e}^{\nu/2}
			\right] 
	= \Delta \Gamma_j
,
\end{gather}
where the reaction rates $\Delta \Gamma_j$ are expressed through
chemical potential imbalances
$\Delta\mu_{e} \equiv \mu_{n} - \mu_{p} - \mu_{e}$
and $\Delta\mu_{\mu} \equiv \mu_{n} - \mu_{p} - \mu_{\mu}$
as
\begin{gather}
\label{eq:gr:DeltaGamma}
	\Delta \Gamma_{n}
	= - \lambda_{e} \Delta\mu_{e}
	  - \lambda_{\mu} \Delta\mu_{\mu}
,\quad
	\Delta \Gamma_{p}
	= \lambda_{e} \Delta\mu_{e}
	  + \lambda_{\mu} \Delta\mu_{\mu}
,\quad
	\Delta \Gamma_{e}
	= \lambda_{e} \Delta\mu_{e}
,\quad
	\Delta \Gamma_{\mu}
	=  \lambda_{\mu} \Delta\mu_{\mu}
,
\end{gather}
and the reaction coefficients $\lambda_{e}$ and $\lambda_{\mu}$
for direct and/or modified Urca processes
can be found, e.g., in Ref.~\cite{ykgh01}.

Energy and momentum conservation laws \eqref{eq:dTmunu=0},
with the help of thermodynamic relations \eqref{eq:2ndlaw} and \eqref{eq:dP},
as well as Maxwell equations, can be presented as:
\begin{gather}
\label{eq:gr:energy-conservation}
	\left(
		\mu_j \pd{n_j}{\hat{t}}
		+ T \pd{S}{\hat{t}}
	\right)
	- {\pmb E} {\pmb J}
	+ {\rm e}^{-\nu} \hat{\pmb \nabla} \left[
		{\rm e}^{\nu} (\mu_j n_j + TS)
		{\pmb u}
	\right]
	= - Q
,\\
\label{eq:gr:momentum-conservation}
	\pd{}{\hat{t}}
	\left[
		(\mu_j n_j+ TS) {\pmb u}
	\right]
	+ n_j {\rm e}^{-\nu/2} 
		\hat{\pmb \nabla}  \left( \mu_j {\rm e}^{\nu/2} \right)
	+ S {\rm e}^{-\nu/2} 
		\hat{\pmb \nabla} \left( T {\rm e}^{\nu/2} \right)
	- e_j n_j {\pmb E} - \left[{\pmb J} \times {\pmb B} \right]
	= 0
.
\end{gather}
In the case of hydrostatic equilibrium Eq.~\eqref{eq:gr:momentum-conservation}
reduces to
\begin{gather}
\label{eq:gr:hydrostatic}
	\hat{\pmb \nabla}  P
	+ (P+\varepsilon) \hat{\pmb \nabla} \frac{\nu}{2}
	= 0.
\end{gather}
One can also write the energy conservation law \eqref{eq:gr:energy-conservation}
in a form of entropy generation equation \eqref{eq:dSmu-EM-LL},
which, with the help of relations \eqref{eq:muDeltaGamma} and \eqref{eq:dj-comoving2},
yields
\begin{gather}
\label{eq:gr:dSmu}
	 \pd{S}{\hat{t}}
	   + {\rm e}^{-\nu/2} ~
	   \hat{\pmb \nabla}
	   \left[
	    \left( S {\pmb u} 
	                           - \frac{\mu_j}{T} \Delta {\pmb j}_{(j)}
	                         \right)
	    {\rm e}^{\nu/2}
	   \right] 
	= \mathcal{D}_{jk}^{\parallel} {\pmb d}_{(j)\parallel} {\pmb d}_{(k)\parallel}
	 + \mathcal{D}_{jk}^{\perp} {\pmb d}_{(j)\perp} {\pmb d}_{(k)\perp}
	+ \frac{\lambda_{e} (\Delta\mu_{e})^2}{T}
	+ \frac{\lambda_{\mu} (\Delta\mu_{\mu})^2}{T}
	- \frac{Q}{T}
.
\end{gather}

Diffusion currents are expressed algebraically through the vectors ${\pmb d}_{(j)}$ [see Eq.~(\ref{eq:djmu2})]:
\begin{gather}
	{\Delta \pmb j}_{(j)}
	= - \mathcal{D}_{jk}^{\parallel} {\pmb d}_{(k)\parallel}
	- \mathcal{D}_{jk}^{\perp} {\pmb d}_{(k)\perp}
	- \mathcal{D}_{jk}^{H} \left[ {\pmb d}_{(k)\perp} \times {\pmb b} \right]
,
\end{gather}
where
\begin{gather}
	{\pmb d}_{(j)} \equiv
	 \hat{\pmb \nabla} \left(\frac{\mu_j}{T}\right) 
	     - \frac{e_j {\pmb E} 
	              + e_j \left[ {\pmb u} \times {\pmb B} \right] 
	            }{T}
.
\end{gather}

To sum up, the MHD equations
contain three unknown vector functions (${\pmb u}$, ${\pmb E}$, ${\pmb B}$)
and five unknown scalars, e.g., $n_{n}$, $n_{p}$, $n_{e}$, $n_{\mu}$, and $S$
(all thermodynamic quantities can be expressed as functions
of $n_{n}$, $n_{p}$, $n_{e}$, $n_{\mu}$, $S$, $|{\pmb B}|^2$, and $|{\pmb E}|^2$, provided the equation of state is specified).

%%%%%%%%%%%%%%%%%%%%%%%%%%%%%%%%%%%%%%%%%%%%%%%%%%%%%%%%%%%%%%%%%%%%
\section{Applications and special cases}
\label{sec:special-cases}
%%%%%%%%%%%%%%%%%%%%%%%%%%%%%%%%%%%%%%%%%%%%%%%%%%%%%%%%%%%%%%%%%%%%

In this section we apply the general hydrodynamic equations presented above
to a number of special cases,
and provide relations between our generalized diffusion coefficients $\mathcal{D}_{jk}$
and various kinetic coefficients commonly used in the literature,
such as electrical conductivity, thermal conductivity, and momentum transfer rates.

\subsection{Electrical conductivity}
\label{sec:el-cond}

In this section we provide relations between
the electrical conductivity and generalized diffusion coefficients $\mathcal{D}_{jk}$.
Let us consider the quasineutral ($e_j n_j = 0$) homogeneous (${\pmb \nabla} \frac{\mu_j}{T} = 0$) matter.
Substituting Eqs. \eqref{eq:dmu} and \eqref{eq:dj-comoving2} into Eq.~\eqref{eq:Jfree},
one obtains the following expression for Ohm's law in the comoving frame:
\begin{gather}
\label{eq:ohm-1}
	{\pmb J}
	= e_j {\Delta \pmb j}_{(j)}
	= \frac{e_j e_k \mathcal{D}_{jk}^{\parallel}}{T} {\pmb E}_{\parallel}
	+ \frac{e_j e_k \mathcal{D}_{jk}^{\perp}}{T} {\pmb E}_{\perp}
	+ \frac{e_j e_k \mathcal{D}_{jk}^{H}}{T} \left[ {\pmb E} \times {\pmb b} \right]
,
\end{gather}
where
${\pmb E}_{\parallel} \equiv ({\pmb E} {\pmb b}) {\pmb b}$
and
${\pmb E}_{\perp} \equiv {\pmb E} - ({\pmb E}_{} {\pmb b}) {\pmb b}$.

Now, introducing
conductivities
$\sigma^\parallel$, $\sigma^\perp$, and $\sigma^H$ 
(describing
the conductivity along ${\pmb B}$,
conductivity in the direction perpendicular to ${\pmb B}$,
and the Hall effect, respectively) 
as
\begin{gather}
\sigma^\parallel = \frac{e_j e_k \mathcal{D}_{jk}^{\parallel}}{T}
,\quad
\sigma^\perp = \frac{e_j e_k \mathcal{D}_{jk}^{\perp}}{T}
,\quad
\sigma^H = \frac{e_j e_k \mathcal{D}_{jk}^{H}}{T}
\end{gather}
and choosing a coordinate frame with the $z$ axis along ${\pmb b}$,
one can easily rewrite
Ohm's law \eqref{eq:ohm-1}
in the following standard form (e.g., Ref.~\cite{ys91a}),
\begin{gather}
\label{eq:ohm-2}
\left(
\begin{array}{c}
 {J}^x
 \\
  {J}^y
 \\
  {J}^z
\end{array}
\right)
=
\left(
\begin{array}{ccc}
 \sigma^\perp & \sigma^H & 0 
 \\
 -\sigma^H & \sigma^\perp & 0
 \\
 0 & 0 & \sigma^\parallel
\end{array}
\right)
\left(
\begin{array}{c}
 E^x
 \\
 E^y
 \\
 E^z
\end{array}
\right)
.
\end{gather}
In the absence of a magnetic field $\sigma^\perp = \sigma^\parallel \equiv \sigma$ and $\sigma^H = 0$,
so that Ohm's law takes the simple form
${\pmb J} = \sigma {\pmb E}$.

\subsection{Thermal conductivity}
\label{sec:thermal}

Let us consider heat conduction in a one-component neutral liquid
without the magnetic field.
In this case the dissipative correction ${\Delta j}_{(1)}^\mu$
\eqref{eq:dj-noB}
takes the form
\begin{gather}
\label{eq:kappa:dj-1}
	{\Delta j}_{(1)}^\mu
	= - \mathcal{D}_{11} \nablaperp^\mu \frac{\mu_1}{T}
.
\end{gather}
Using Eq.~\eqref{eq:pres}, the Gibbs-Duhem relation \eqref{eq:dP}
is presented as
\begin{gather}
\label{eq:kappa:dP}
	dP = n_1 T \, d\frac{\mu_1}{T} + (P+\varepsilon) \, \frac{dT}{T}
.
\end{gather}
Using this relation
and introducing the thermal conductivity coefficient\footnote{
In the presence of a magnetic field one can introduce, in analogy with the electrical conductivity,
	the quantities
	$\kappa^\parallel \equiv \mathcal{D}_{11}^\parallel \left( \frac{P+\varepsilon}{n_1 T} \right)^2$,
	$\kappa^\perp     \equiv \mathcal{D}_{11}^\perp     \left( \frac{P+\varepsilon}{n_1 T} \right)^2$, and
	$\kappa^H         \equiv \mathcal{D}_{11}^H         \left( \frac{P+\varepsilon}{n_1 T} \right)^2$.
}
\begin{gather}
\label{eq:kappa:kappa}
	\kappa \equiv \mathcal{D}_{11} \left( \frac{P+\varepsilon}{n_1 T} \right)^2
,
\end{gather}
Eq.~\eqref{eq:kappa:dj-1}
yields (see Ref.~\cite{ll87}, Sec.~139)
\begin{gather}
\label{eq:kappa:dj-2}
	{\Delta j}_{(1)}^\mu
	= \kappa \frac{n_1}{P+\varepsilon}
		\left(
			\nablaperp^\mu T 
			- \frac{T}{P+\varepsilon} \nablaperp^\mu P			
		\right)
.
\end{gather}
Now let us define the particle four-velocity
$V^\mu$:
\begin{gather}
\label{eq:kappa:Vmu}
	V^\mu \equiv j_{(1)}^\mu / n_1 =  u^\mu + \frac{1}{n_1} {\Delta j}_{(1)}^\mu
,
\end{gather}
normalized by the condition $V_{\mu}V^{\mu}=-1$, valid to linear order in small dissipative corrections 
[see Eq.\ (\ref{eq:condjmu2})].
In the ``particle frame'', defined by condition $V^\mu = (1,0,0,0)$, 
the energy density four-current reads\footnote{
We omit the quadratically small term $-V_\nu \Delta \tau^{\mu\nu} = \Delta j_{(1)\nu} \Delta \tau^{\mu\nu} / n_1$.
}
\begin{gather}
	 -V_\nu T^{\mu\nu}
	 = \varepsilon V^\mu
	   - \kappa 
	 		\left(
	 			\nablaperp^\mu T 
	 			- \frac{T}{P+\varepsilon} \nablaperp^\mu P			
	 		\right)
.
\end{gather}
The entropy four-current [see Eq.~\eqref{eq:dSmu-EM-LL-0}]
can be expressed, 
with the help of Eqs.~\eqref{eq:kappa:dj-2} and \eqref{eq:kappa:Vmu},
as
\begin{gather}
\label{eq:kappa:Smu}
	S^\mu
	= S V^\mu
	- \frac{\kappa }{T} 
	 		\left[
	 			\nablaperp^\mu T 
	 			- \frac{T}{P+\varepsilon} \nablaperp^\mu P			
	 		\right]
.
\end{gather}

Note that the same consideration remains valid,
e.g., for multicomponent nonsuperfluid $npe\mu$ matter in NS cores,
if one assumes that beta processes are frozen and
all particles move with the same velocity $V^\mu$,
so that the system can be treated as the single-component one.
Let us consider thermal evolution of a spherically symmetric NS under this assumption,
ignoring particle currents [$V^\mu = (1,0,0,0)$ in the laboratory frame], the magnetic field and nonequilibrium reactions
but taking into account effects of general relativity, described in Sec.~\ref{sec:gr}.
Using the hydrostatic equilibrium condition \eqref{eq:gr:hydrostatic},
one can rewrite the combination
$\hat{\pmb \nabla} T - \frac{T}{P+\varepsilon} \hat{\pmb \nabla} P$
as
\begin{gather}
\label{eq:kappa:dT}
	\hat{\pmb \nabla} T 
	- \frac{T}{P+\varepsilon} \hat{\pmb \nabla} P
	= 	{\rm e}^{-\nu/2}~\hat{\pmb \nabla} \left(T {\rm e}^{\nu/2}\right) 
.
\end{gather}
In view of Eqs.~\eqref{eq:kappa:kappa}, \eqref{eq:kappa:dj-2}, \eqref{eq:kappa:Smu} and \eqref{eq:kappa:dT}
the entropy generation equation \eqref{eq:gr:dSmu}
reduces to
\begin{gather}
	 \pd{S}{\hat{t}}
	 - {\rm e}^{-\nu/2} ~
	 \hat{\pmb \nabla}
	 \left[
	 	 \frac{\kappa}{T}
		  \hat{\pmb \nabla} \left(T {\rm e}^{\nu/2}\right) 
	 \right] 
	 = \kappa \left|
		 	\frac{\hat{\pmb \nabla} \left(T {\rm e}^{\nu/2}\right)}
			 	{T {\rm e}^{\nu/2}}
	 	\right|^2
		- \frac{Q}{T}
%,
\end{gather}
or, equivalently,
\begin{gather}
	C_v \pd{T}{\hat{t}}
	- {\rm e}^{-\nu} \hat{\pmb \nabla}
	 \left[
	 	\kappa {\rm e}^{\nu/2} ~
		  \hat{\pmb \nabla} \left(T {\rm e}^{\nu/2}\right)
	 \right] 	
	 = - Q
,
\label{eqQ}
\end{gather}
where
$C_v \equiv T \partial S/ \partial T$
is the heat capacity per unit volume.
Using the definitions of $\pd{}{\hat{t}}$ \eqref{eq:gr:dt-hat}
and $\hat{\pmb \nabla}$ \eqref{eq:gr:nabla-hat},
one finally represents Eq.\ (\ref{eqQ}) in
the following standard form (see, e.g., Ref.\ \cite{ykgh01}):
\begin{gather}
\label{eq:kappa:thermal-final}
	{\rm e}^{-\nu -\lambda/2}
	\frac{1}{4\pi r^2} 
		\pd{}{r} \left( {\rm e}^{\nu} L_r \right)
	= -Q - 	C_v {\rm e}^{-\nu/2} \pd{T}{t}
,\\
\label{eq:kappa:Lr}
	\frac{L_r}{4\pi r^2}
	= - \kappa	{\rm e}^{-(\nu+\lambda)/2}
		\pd{}{r} \left(T {\rm e}^{\nu/2}\right)
,
\end{gather}
where the ``local luminosity'' $L_r$ is the (not related to neutrinos) heat flux transported through a sphere of radius $r$.

We remind the reader that Eqs.~\eqref{eq:kappa:thermal-final} and \eqref{eq:kappa:Lr}
are valid only in the absence of particle currents and deviations from beta equilibrium;
to study thermal evolution
under more realistic assumptions,
one has to use a more general equation \eqref{eq:gr:dSmu}.

\subsection{Diffusion, thermodiffusion and thermal conductivity in the nonrelativistic limit}
\label{sec:nonrel}

In this section, we derive 
a relation between our diffusion coefficients
and standard coefficients
of diffusion, thermodiffusion, and thermal conductivity, arising
in the nonrelativistic hydrodynamics.
In the nonrelativistic limit
the relativistic chemical potential $\mu_j$ for particle species $j$
approximately coincides with its rest mass energy, $m_j c^2$,
so that generally
$\left| {\pmb \nabla}\mu_j / T \right| 
\ll \left| \mu_j {\pmb \nabla} \left(1/ T\right) \right|
\approx  \left| m_j c^2 {\pmb \nabla} \left(1/ T\right) \right|
$,
and, thus, at arbitrary $\mathcal{D}_{jk}$ the terms depending on gradients of temperature will be dominant in the expressions
\eqref{eq:dj-noB} and \eqref{eq:djmu2}
for $\Delta j_{(j)}^\mu$.
However, we know that in the nonrelativistic hydrodynamics
both ``chemical potential'' and ``temperature''
terms can be equally important.
To resolve the seeming contradiction, as we show below, one has to impose additional constraints on the coefficients in the nonrelativistic expansion of $\mathcal{D}_{jk}$.
For the sake of simplicity,
we consider a neutral binary mixture.
We also ignore viscosity and chemical reactions, since these effects do not interfere with diffusion.
In this section (as well as in Secs. \ref{sec:comparison} and \ref{sec:Jik} and in Appendix~\ref{sec:ncomp})
we do not set $c=1$ to make the transition to the nonrelativistic limit more transparent.

Let us expand 
$\mu_j$ and $\mathcal{D}_{jk}$
in small parameter $\delta = v^2/c^2$,
where $v$ is a typical microscopic particle velocity in the mixture:
\begin{gather}
\label{eq:nonrel:mu-series}
	\mu_j = m_j c^2 + \mu_j^{(1)} \delta
	+ \mu_j^{(2)} \delta^2 
	+ O(\delta^3)
,\\
\label{eq:nonrel:D-series}
	\mathcal{D}_{jk}
	= \mathcal{D}_{jk}^{(0)}
	+ \mathcal{D}_{jk}^{(1)} \delta
	+ \mathcal{D}_{jk}^{(2)} \delta^2
	+ O(\delta^3)
.
\end{gather}
In view of Eqs.~\eqref{eq:dmu}, \eqref{eq:nonrel:mu-series}, and \eqref{eq:nonrel:D-series},
the dissipative corrections to particle currents [Eq.~\eqref{eq:dj-noB}] can be expanded in the comoving frame 
[$\Delta j_{(j)}^\mu = (0, \Delta {\pmb j}_{(j)})$] as
\begin{gather}
\label{eq:nonrel:dj-series}
	\Delta {\pmb j}_{(j)}
	= - \mathcal{D}_{jk}^{(0)} m_k c^2 {\pmb \nabla} \frac{1}{T}
	  - \left(
		  	\mathcal{D}_{jk}^{(0)} {\pmb \nabla} \frac{\mu_k^{(1)}}{T}
		  	+ \mathcal{D}_{jk}^{(1)} m_k c^2 {\pmb \nabla} \frac{1}{T}
		  \right)
		  \delta 
	  - \left(
	        \mathcal{D}_{jk}^{(0)} {\pmb \nabla} \frac{\mu_k^{(2)}}{T}
		  	+ \mathcal{D}_{jk}^{(1)} {\pmb \nabla} \frac{\mu_k^{(1)}}{T}
		  	+ \mathcal{D}_{jk}^{(2)} m_k c^2 {\pmb \nabla} \frac{1}{T}
		 \right)
		 \delta^2 
	+ O(\delta^3)
.
\end{gather}

In the nonrelativistic theory (see, e.g., Sec.~59 in Ref.\ \cite{ll87}),
the effects of diffusion, thermodiffusion and thermal conductivity in a binary mixture
are described in terms of the heat current ${\pmb q}$
and the diffusion current ${\pmb i}$, 
which can be expressed in terms of $\Delta {\pmb j}_{(j)}$
as\footnote{
To obtain the expression for ${\pmb q}$
one should express the components of the energy-momentum tensor $T^{0i}$ ($i=1,2,3$)
through $\Delta j_{(j)}^\mu$ and the center-of-mass velocity
$V^\mu \equiv u^\mu + m_j c \Delta j_{(j)}^\mu /\rho$ ($\rho$ is the total mass density measured in the laboratory frame)
and compare the result with the nonrelativistic expression for $T^{0i}$
\cite{ll87}.
}
\begin{gather}
\label{eq:nonrel:i}
	{\pmb i}
	= \frac{m_1 n_1 m_2 n_2}{m_1 n_1 + m_2 n_2}
		\left(
			\frac{\Delta {\pmb j}_{(1)}}{n_1}
		  - \frac{\Delta {\pmb j}_{(2)}}{n_2}
		\right)
,\\
\label{eq:nonrel:q}
	{\pmb q} = - m_1 c^2 \Delta {\pmb j}_{(1)}
	           - m_2 c^2 \Delta {\pmb j}_{(2)}
,
\end{gather}
respectively,
where we label the components of the mixture by indices $1$ and $2$.

The currents ${\pmb i}$ and ${\pmb q}$
depend on gradients of temperature $T$
and chemical potential
$\mu_{\rm LL}
\equiv \mu_1 / m_1 - \mu_2 / m_2
= \left( \mu_1^{(1)} / m_1 - \mu_2^{(1)} / m_2 \right) \delta
+ O(\delta^2)
$
through the nonrelativistic kinetic coefficients $\alpha$, $\beta$, and $\gamma$
\cite{ll87}:
\begin{gather}
	{\pmb i}
	= - \alpha {\pmb \nabla} \mu_{\rm LL} - \beta {\pmb \nabla} T
,\\
	{\pmb q}	
	= - \beta T {\pmb \nabla} \mu_{\rm LL} - \gamma {\pmb \nabla} T + \mu_{\rm LL} {\pmb i}
.
\end{gather}

In order to relate $\alpha$, $\beta$, and $\gamma$ with the generalized diffusion coefficients $\mathcal{D}_{jk}$,
one has to substitute the expansion~\eqref{eq:nonrel:dj-series}
into relations  \eqref{eq:nonrel:i} and \eqref{eq:nonrel:q},
retaining the lowest-order terms in $\delta$.
In order to reproduce the nonrelativistic results,
where ${\pmb \nabla} T$ and ${\pmb \nabla} \mu_j$
enter the expressions for ${\pmb i}$ and ${\pmb q}$ on an equal footing,
we require the first term in Eq.~\eqref{eq:nonrel:dj-series}
to vanish, or, equivalently,\footnote{
Note that a similar condition \eqref{eq:muD=0}
holds also for degenerate (even relativistic) matter,
if we expand $\mathcal{D}_{jk}$ in powers of $\delta = O \left( T/\mu_j \right)$.
}
\begin{gather}
\label{eq:nonrel:mD0=0}
	m_k \mathcal{D}_{jk}^{(0)} = 0
.
\end{gather}
Using this relation and calculating the heat current ${\pmb q}$ to the first order in $\delta$, we find
\begin{gather}
\label{eq:nonrel:q-1}
	{\pmb q}
	= \left[ m_1^2 \mathcal{D}_{11}^{(1)} 
		  + 2 m_1 m_2 \mathcal{D}_{12}^{(1)} 
		  + m_2^2 \mathcal{D}_{22}^{(1)} 
		\right]
		c^4 {\pmb \nabla} \left( \frac{1}{T} \right)
		\delta 
	+ O(\delta^2)
.
\end{gather}
This means that ${\pmb q}$ is, generally, independent of gradients of chemical potentials in the limit $\delta \to 0$.
Thus, to reproduce the nonrelativistic results, one has to require, additionally,
\begin{gather}
\label{eq:nonrel:mmD1=0}
	m_1^2 \mathcal{D}_{11}^{(1)} 
	+ 2 m_1 m_2 \mathcal{D}_{12}^{(1)} 
	+ m_2^2 \mathcal{D}_{22}^{(1)}
	= 0
,
\end{gather} 
so that ${\pmb q} = O(\delta^2)$ [but both ${\Delta {\pmb j}_{(j)}}$ and ${\pmb i}$ are $O(\delta)$].
Excluding $\mathcal{D}_{11}^{(0)}$, $\mathcal{D}_{22}^{(0)}$, and $\mathcal{D}_{12}^{(1)}$
via relations
\eqref{eq:nonrel:mD0=0} and \eqref{eq:nonrel:mmD1=0},
one finally arrives at the following expressions for the nonrelativistic coefficients,
\begin{gather}
\label{eq:nonrel:a}
	\alpha
	= - \frac{m_1 m_2 \mathcal{D}_{12}^{(0)} }{T} 
,\\
\label{eq:nonrel:b}
	\beta
	= \frac{1}{2 T^2}
	\left[
		2 m_2 m_1 \mathcal{D}_{12}^{(0)} \mu_{\rm LL} 
		- m_1^2 c^2 \mathcal{D}_{11}^{(1)} \delta 
		+ m_2^2 c^2 \mathcal{D}_{22}^{(1)} \delta 
	\right]
,\\
\label{eq:nonrel:g}
	\gamma
	= \frac{1}{T^2}
		\left[
			m_1 m_2
			\left(
				2 c^4 \mathcal{D}_{12}^{(2)} \delta^2 - \mathcal{D}_{12}^{(0)} \mu_{\rm LL}^2
			\right)
			+m_1^2 c^2
			\left(
				c^2 \mathcal{D}_{11}^{(2)} \delta^2 + \mathcal{D}_{11}^{(1)} \mu_{\rm LL} \delta
			\right)
			+ m_2^2 c^2
			\left(
				c^2 \mathcal{D}_{22}^{(2)} \delta^2 - \mathcal{D}_{22}^{(1)} \mu_{\rm LL} \delta
			\right)
		\right]
.
\end{gather}

To sum up, we expressed the nonrelativistic kinetic coefficients $\alpha$, $\beta$, and $\gamma$,
which describe diffusion, thermodiffusion, and thermal conductivity,
through coefficients in the nonrelativistic expansion
of $\mathcal{D}_{jk}$ \eqref{eq:nonrel:D-series}.
In addition, we obtained two constraints on these expansion coefficients
in the zeroth \eqref{eq:nonrel:mD0=0} and first \eqref{eq:nonrel:mmD1=0} order in the expansion parameter $\delta$.

%%%%%%%%%%%%%%%%%%%%%%%%%%%%%%%%%%%%%%%%%%%%%%%%%%%%%%%%%%%%%%%%%%%%
%\section{Comparison with results of Refs.\ \cite{ys91a,gr92}}
\subsection{Momentum transfer rates and diffusion in the low-temperature limit}
\label{sec:comparison}
%%%%%%%%%%%%%%%%%%%%%%%%%%%%%%%%%%%%%%%%%%%%%%%%%%%%%%%%%%%%%%%%%%%%

In this section we compare our approach with the microscopic formalism
used, e.g., in Refs.~\cite{ys91a,gr92,papm17,gko17,og18} in the limit $T \to 0$
and assuming that ${\pmb \nabla} T = 0$. We express our generalized diffusion coefficients
$\mathcal{D}_{jk}^\parallel$, $\mathcal{D}_{jk}^\perp$, and $\mathcal{D}_{jk}^H$
through the momentum transfer rates $J_{ik}$ introduced in the microscopic theory.

The general multicomponent equations describing nonsuperfluid liquid
are similar to those used in Ref.\ \cite{gr92} (see also \cite{ys91a}
for analogous equations).
Let us  assume that the liquid constituents move with the nonrelativistic velocities ${\pmb u}_j$ (the equation of state for the liquid can be nevertheless relativistic). 
In the hydrodynamic regime investigated throughout the paper, the velocities ${\pmb u}_j$ almost coincide due to frequent collisions
(e.g., \cite{brag65}),
and it is always possible to choose the frame where ${\pmb u}_j \ll c$.
In this case, the Euler equation for particle species $j$ reads %
(hereafter in this section no summation over particle indices $j,k,\ldots$ is assumed)
\begin{gather}
\label{eq:ys1}
	n_{j}
	\left[
		\frac{\partial}{\partial t}
		+ ({\pmb u}_j {\pmb \nabla})
	\right]
	\left(
		\frac{\mu_{j}}{c^2} {\pmb u}_j
	\right)
	=
	e_{j} n_{j} \left( {\pmb E}+\frac{1}{c}\left[{\pmb u}_{j} \times {\pmb B} \right]\right)
	-n_{j} {\pmb \nabla}\mu_{j} 
	- \frac{\mu_{j} n_{j}}{c^2} \, {\pmb \nabla}\phi
	-\sum_{k \neq j} J_{jk}({\pmb u}_{j}-{\pmb u}_{k})
,
\end{gather}
where ${\pmb u}_j$ is the velocity for particle species $j$;
$c$ is the speed of light; 
$\phi$ is the gravitational potential;
and $J_{jk} = J_{kj}$ is the momentum transfer rate between particle species $j$ and $k$
per unit volume, which is related to the effective relaxation time $\tau_{jk}$
by the formula $J_{jk} =\mu_{j}n_{j}/(c^2\tau_{jk})$.

In the hydrodynamic regime, when collision timescales
are much smaller than the typical hydrodynamic timescale, 
velocities ${\pmb u}_{j}$ are very close to one another \cite{brag65}
and one can replace
in lhs of Eq.~\eqref{eq:ys1} ${\pmb u}_{j}$ with the average mass velocity ${\pmb U}$,
defined as \cite{ys91a}%
%see gko17 after eq 11:
%
\footnote{
For example, the term
$ n_j \partial / \partial t
	\left[
		\mu_{j}  \left({\pmb u}_j - {\pmb U} \right) / c^2
	\right]
\sim
    \mu_j n_j \left({\pmb u}_j - {\pmb U} \right) / (\mathcal{T} c^2)
$,
where $\mathcal{T}$ is a typical timescale of the problem,
can be neglected in comparison to the term
$\sum_{k \neq j} J_{jk}({\pmb u}_{j}-{\pmb u}_{k})
\sim \sum_{k \neq j} \mu_{j}n_{j}  ({\pmb u}_{j}-{\pmb u}_{k}) / (c^2\tau_{jk})
$, because $\mathcal{T}\gg \tau_{jk}$.
}
\begin{gather}
\label{eq:ys2}
	{\pmb U} \sum_{j} \mu_{j} n_{j} \equiv     
	\sum_{j} \mu_{j}n_{j} \, {\pmb u}_{j}.
\end{gather}
After the replacement, Eq.~\eqref{eq:ys1} becomes
\begin{gather}
\label{eq:ys3}
	n_{j}
	\left[
		\frac{\partial}{\partial t}
		+ ({\pmb U} {\pmb \nabla})
	\right]
	\left(
		\frac{\mu_{j}}{c^2} {\pmb U}
	\right)	
	=
	e_{j} n_{j} \left( {\pmb E}+\frac{1}{c}\left[{\pmb u}_{j} \times {\pmb B} \right]\right)
	-n_{j} {\pmb \nabla}\mu_{j} 
	- \frac{\mu_{j} n_{j}}{c^2} \, {\pmb \nabla}\phi
	-\sum_{k \neq j} J_{jk}({\pmb u}_{j}-{\pmb u}_{k})
.
\end{gather}
One needs one more equation to specify the frame in which Eq.~\eqref{eq:ys3} is written;
as in Ref.\  \cite{ys91a}, we define it by the condition ${\pmb U} = 0$ (at a given moment of time)
or, equivalently,
\begin{gather}
\label{eq:mnu=0}
	\sum_j \mu_j n_j {\pmb u}_j = 0
.
\end{gather}
In this frame Eq.~\eqref{eq:ys3} reduces to
\begin{gather}
\label{eq:ys4}
	\frac{\mu_j n_j}{c^2}
		\frac{\partial {\pmb U}}{\partial t}
	=
	e_{j} n_{j} \left( {\pmb E}+\frac{1}{c}\left[{\pmb u}_{j} \times {\pmb B} \right]\right)
	-n_{j} {\pmb \nabla}\mu_{j} 
	- \frac{\mu_{j} n_{j}}{c^2} \, {\pmb \nabla}\phi
	-\sum_{k \neq j} J_{jk}({\pmb u}_{j}-{\pmb u}_{k})
.
\end{gather}
Excluding plasma acceleration from Eq.~\eqref{eq:ys4}, one obtains,
with the aid of Eq.~\eqref{eq:mnu=0},
the velocities ${\pmb u}_j$ for each particle species as algebraic functions of thermodynamic forces.

The total energy current density in the microscopic formalism \cite{ys91a,gr92} reads
\begin{gather}
\label{eq:T0i-ys}
	{\pmb q}
	= \sum_j \mu_j n_j {\pmb u}_j
	+ \frac{c}{4\pi} [{\pmb E} \times {\pmb B}]
.
\end{gather}
In view of Eq.~\eqref{eq:mnu=0}, the first term in \eqref{eq:T0i-ys} vanishes in the ${\pmb U}=0$ frame.

Let us now consider the same situation in our formalism. The total energy current density ${\pmb q}$,
determined by the components of the energy-momentum tensor as $q^i = c T^{0i}$ ($i=1,2,3$), 
in the comoving frame reads
\begin{gather}
\label{eq:T0i}
	q^i
	= c \Delta T_{({\rm EM})}^{0i}
	= \frac{c}{4\pi} [{\pmb E} \times {\pmb B}]^i
.
\end{gather}
Clearly, comparing Eqs.~\eqref{eq:T0i} and \eqref{eq:T0i-ys},
one can see that the comoving frame $u^\mu = (1,0,0,0)$
coincides with the frame defined by the condition~\eqref{eq:mnu=0}.

Solving the system~\eqref{eq:mnu=0} and \eqref{eq:ys4},
one can express velocities ${\pmb u}_j$, or, equivalently,
particle currents $\Delta {\pmb j}_{(j)} = n_j {\pmb u}_j /c$
through the vectors
${\pmb d}_{(j)} = \frac{{\pmb \nabla}\mu_j - e_j {\pmb E}}{T}$
[cf. definition~\eqref{eq:dmu} with ${\pmb \nabla} T = 0$],
magnetic field ${\pmb B}$,
momentum transfer rates $J_{jk}$,
and the equilibrium thermodynamic quantities.
Comparing the result with Eq.~\eqref{eq:dj-comoving2},
one can translate one formalism into another
and find diffusion coefficients
$\mathcal{D}_{jk}^\parallel$, $\mathcal{D}_{jk}^\perp$, and $\mathcal{D}_{jk}^H$.
The general algorithm is presented in Appendix~\ref{sec:ncomp}.
In the simplest case of a binary mixture in the absence of a magnetic field
one obtains
\begin{gather}
	\mathcal{D}_{12}
	= \mathcal{D}_{21}
	= - \frac{n_1^2 n_2^2 T \mu_1 \mu_2}
			{J_{12} (\mu_1 n_1 + \mu_2 n_2)^2 c} 
,\\
\label{eq:muD=0}
	\mathcal{D}_{11} = - \frac{\mu_2}{\mu_1} \mathcal{D}_{12}
,\quad
	\mathcal{D}_{22} = - \frac{\mu_1}{\mu_2} \mathcal{D}_{12}
.
\end{gather}
We present the momentum transfer rates $J_{ik}$
for $npe\mu$ matter in Sec.~\ref{sec:Jik}.

%%%%%%%%%%%%%%%%%%%%%%%%%%%%%%%%%%%%%%%%%%%%%%%%%%%%%%%%%%%%%%%%%%%%%%%%%%%%%
\section{Momentum transfer rates in $npe\mu$ matter}
\label{sec:Jik}
%%%%%%%%%%%%%%%%%%%%%%%%%%%%%%%%%%%%%%%%%%%%%%%%%%%%%%%%%%%%%%%%%%%%%%%%%%%%%

Bearing in mind an application of our results to the problem of magnetic field evolution in an NS core, here we present the practical expressions for the momentum transfer rates $J_{ik}$,
introduced in the previous section,
for the simplest $npe\mu$ composition.
Here we do not set $c=k_{\rm B}=1$.
In most of the present studies, the microscopic calculations under the ``free particle model'' from Ref.~\cite{ys91b} are adopted
(see, e.g., Refs.\ \cite{gr92,papm17}). 
However, there are considerable updates to their results in the past decades;
see Ref.~\cite{SchmittShternin2018} for a review. 

The collisions in NS matter are divided in two sectors. The first one includes collisions between the leptons and charged baryons, $e\mu$, $ep$, and $\mu p$ in the present case, and  is governed by the electromagnetic interactions. The second sector contains the baryon collisions, $np$ in the present case, which are mediated by the strong interactions. We remind the reader that the like-species collisions do not contribute to the diffusion rates. 

The momentum transfer rates for the electromagnetic collisions should be calculated taking into account the correct plasma screening \cite{SchmittShternin2018}. Appropriate expressions have been derived in Ref.~\cite{shternin08}. The momentum transfer rate is a sum of two terms which have different temperature dependences reflecting different characters of screening of ``electric'' and ``magnetic'' parts of the interaction:
\begin{equation}\label{eq:Jelectro}
  J_{ik} = J_{ik}^l + J_{ik}^t,
\end{equation}
where $J_{ik}^l\propto T^2$ describe the interaction via
the longitudinal plasmon exchange and the dominant term
$J_{ik}^t\propto T^{5/3}$ corresponds to the exchange of transverse plasmons. In the leading order \cite{shternin08},
\begin{equation}\label{eq:Jt}
J^{t}_{ik}=\frac{4\xi^t}{3\pi^3} e_i^2 e_k^2\frac{p_{Fi}^2p_{Fk}^2}{\hbar^6 c^3} \frac{\left(k_BT\right)^{5/3}}{\left(\hbar c q_t\right)^{2/3}},
\end{equation}
\begin{equation}
\label{eq:Jl}
	J^l_{ik} = \frac{4}{9\pi} e_i^2 e_k^2 \frac{m_i^{*2}m_k^{*2} c}{\hbar^6 } \frac{(k_B T)^2} {\hbar c q_l}I_{\ell 2}(q_m/q_l),
\end{equation}
where $p_{Fi}$ and $p_{Fk}$ are the colliding particles Fermi momenta; $m_{i}^*$ and $m^*_{k}$ are their effective masses on the Fermi surface (related to the density of states); for leptons $m_{\ell}\equiv \mu_\ell/c^2$, and $\xi_t=1.813$. Notice that in the $npe\mu$ matter all charged particles have charges $e_i=\pm e$, where $e$ is the elementary electron charge. The quantities $q_l$ and $q_t$ in Eqs.~(\ref{eq:Jt}) and (\ref{eq:Jl}) are the characteristic screening momenta in the plasma:
\begin{eqnarray}
q_l^2&=&\frac{4}{\pi \hbar^3}\sum_i e_i^2 p_{Fi} m_{i}^*,\label{eq:qt}\\
q_t^2&=&\frac{4}{\pi \hbar^3 c}\sum_i e_i^2 p_{Fi}^2.\label{eq:ql}
\end{eqnarray}
The summation in Eqs.~(\ref{eq:qt}) and (\ref{eq:ql}) is carried over all charged components of the plasma, in this sense the collisions between two particle species are influenced by all charged species via the plasma mean field. Finally, 
the function $I_{\ell 2}(q_m/q_l)$ in Eq.~(\ref{eq:Jl}), where $\hbar q_m=2\max(p_{Fk},\,p_{Fi})$, is \cite{shternin08}
\begin{equation}
    I_{\ell 2}(x)=\frac{1}{2} \mathrm{atan}(x) -\frac{1}{2}\frac{x}{1+x^2}.
\end{equation}
If $q_l\ll q_m$, it is enough to take $I_{\ell 2}=\pi/4$; then Eq.~(\ref{eq:Jl}) does not depend on $q_m$. 
The lepton-neutron coupling is small and usually can be neglected. It arises from magnetic interaction with the neutron magnetic moment. For completeness, retaining only the dominant transverse part of the interaction, one obtains
\begin{equation}\label{eq:Jelln}
    J_{\ell n} = \frac{8\pi \alpha_f^2 F_n^2 n_\ell}{9\hbar c^2} (k_B T)^2,
\end{equation}
where $F_n=1.91$ is the neutron magnetic moment in nuclear units, $\ell$ stands for a given lepton (electron or muon), and $\alpha_f=1/137$ is the fine structure constant.

\begin{figure}[th]
    \centering
\includegraphics[width=0.6\columnwidth]{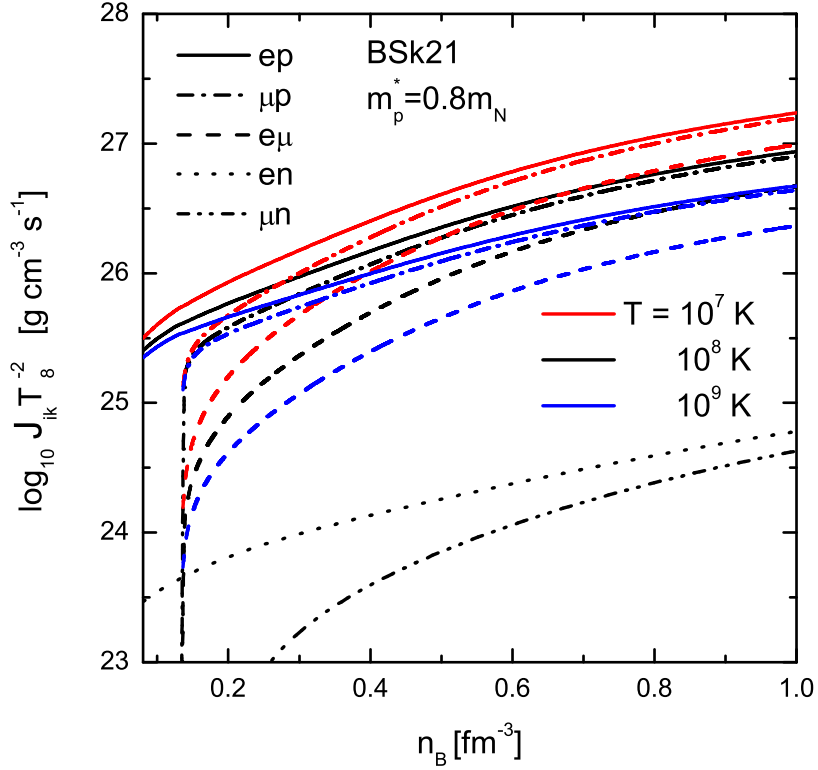}    
\caption{Momentum transfer rates $J_{ik}$ for the electromagnetic sector in the beta-stable matter with the BSk21 EOS.}
    \label{fig:Jemn}
\end{figure}

Equations~(\ref{eq:Jelectro})-(\ref{eq:Jelln}) allow one to calculate the rates of momentum transfer in the electromagnetic sector for any equation of state, provided the effective masses of all charged particles (including baryons) are known. Usually, Eq.~(\ref{eq:Jt}) is enough when the collisions between light relativistic particles (leptons) are considered, while for the massive baryons (protons in our case) both terms in Eq.~(\ref{eq:Jelectro}) need to be preserved due to a large value of baryon mass. For illustration, in Fig.~\ref{fig:Jemn} we plot the momentum transfer rates $J_{ik} T_{8}^{-2}$,
where $T_8\equiv T/(10^8~\mathrm{K})$,  as functions of a baryon density for a beta-stable matter with the  BSk21 EOS, which is based on  the Brussels-Skyrme nucleon interaction functionals \cite{Potekhin2013A&A}. Solid, dashed, and dotted lines show the momentum transfer rates for electron collisions with protons, muons, and neutrons, respectively. Likewise, dot-dashed and double-dot-dashed lines are for the muon-proton and muon-neutron collisions, respectively. In the expressions which depend on the proton effective mass, the latter is set to $m_p^*=0.8m_N$. Because of the dynamical character of plasma screening, for the $e\mu$, $ep$, and $p\mu$ collisions, the combination $J_{ik}T^2$ is not temperature independent. By red, black, and blue lines we show in Fig.~\ref{fig:Jemn} the momentum transfer rates which correspond to $T=10^7$, $10^8$, and $10^9$~K, respectively. One observes that at high densities, the electron-proton and muon-proton momentum transfer rates are close to each other, while the electron-muon momentum transfer rate is several times smaller. The lepton-neutron rates are, in general, significantly smaller than all other rates. Nevertheless, they can be important in the core of a neutron star  under the presence of a strong proton superconductivity (if the neutrons are not in the paired state).

Calculations of the neutron-proton  momentum transfer rate $J_{np}$ are more involved, since this rate depends on the uncertain properties of the nucleon interactions in the dense asymmetric nuclear matter.  \citet{ys91b} used the approximation of free-space zero-angle $np$ cross section (thus neglecting its angular dependence). This approximation is rather crude and leads to a significant overestimate of $J_{np}$ even in the free-space model for the neutron-proton scattering. 
The expression which accurately takes into account the energy and angular dependence of the $np$ scattering cross section can be constructed based on the results of Ref.~\cite{Baiko2001A&A} for the thermal conductivity of NS cores. One obtains \cite{shternin08}
\begin{equation}\label{eq:Jnp}
J_{np}=\frac{64 m_n^{*2} m_p^{*2} (k_BT)^2}{9\pi^2 m_N^2 \hbar^6}
p_{Fn}^3 S_{p2},
\end{equation}
where $m_N$ is the bare nucleon mass and the function $S_{p2}$, having the dimension of a cross section, is defined in Ref.~\cite{Baiko2001A&A}. In that paper the function $S_{p2}$ is calculated and fitted employing accurate free-space differential scattering cross section. The resulting expression is
\begin{widetext}
\begin{equation}\label{eq:SBHY}
S_{p2}=\frac{0.3830k_{Fp}^4}{k_{F
n}^{5.5}} \frac{1+102.0k_{F p}+53.91 k_{F
n}}{1-0.7087k_{F n}+0.2537k_{F n}^2+9.404k_{F
p}^2-1.589k_{F n}k_{F p}}\; \mathrm{mb},
\end{equation}
\end{widetext}
where $k_{Fn}$ and $k_{Fp}$ are neutron and proton Fermi momenta in units of $\mathrm{fm}^{-1}$, respectively, and the fit is valid for the momenta ranges
$k_{Fn}=1.1 - 2.6$ and $k_{Fp}=0.3 - 1.2$.

\begin{figure}[th]
    \centering
\includegraphics[width=0.6\columnwidth]{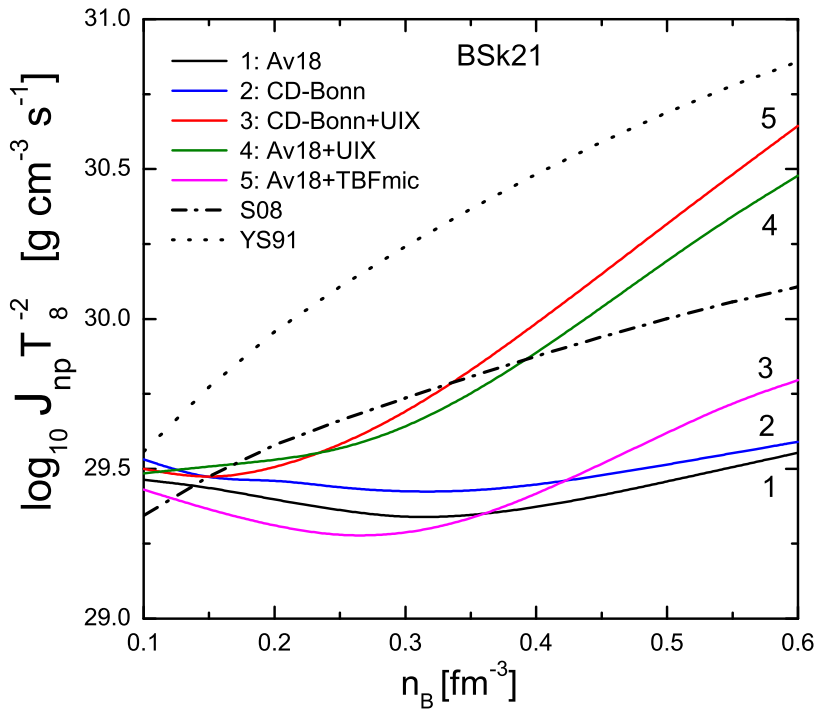}    
\caption{Momentum transfer rates $J_{np}$ in the beta-stable matter with the BSk21 EOS. Solid curves correspond to different nucleon potentials following Ref.~\cite{Shternin2017JPhCS}; number codes are expanded in the legend. The dotted line shows the approximation from Ref.~\cite{ys91b}. The dash-dotted line is calculated following Ref.~\cite{shternin08}, where the fit from Ref.~\cite{Baiko2001A&A} was used. See the text for details.}
    \label{fig:Jnp}
\end{figure}

The expressions (\ref{eq:Jnp}) and (\ref{eq:SBHY}) have the advantage that they can be used for any EOS of the nuclear matter. However, it is known that the in-medium effects considerably modify both the effective masses and scattering cross sections, which in turn modify the transport coefficients; see, e.g., Ref.\ \cite{SchmittShternin2018}, Sec.~IV.A.3, and references therein. Baiko, Heansel, and Yakovlev \cite{Baiko2001A&A} included in-medium corrections [their Eq.~(30)] employing the results of particular Dirac-Brueckner calculations for symmetric nuclear matter ($n_p=n_n$). It is expected, however, that the influence of medium effects on the $np$ cross sections in the asymmetric matter, which is relevant to NS cores, 
can be quantitatively different from the symmetric matter case (see, for instance, Ref.\ \cite{Shternin2013PhRvC}). Unfortunately, it is hard to employ the in-medium corrections in a definite way. They depend on the particular model of the nuclear interaction, the asymmetry of matter, and the many-body theory used. 
Even within a selected many-body approach, for instance, Brueckner-Hartree-Fock, the resulting transport coefficients can differ by an order of magnitude for different models of nuclear interaction \cite{Shternin2017JPhCS}.
We illustrate this uncertainty in Fig.~\ref{fig:Jnp}, where the momentum transfer rates $J_{np} T_{8}^{-2}$
are plotted as a function of the baryon density for the same BSk21 EOS as used in
Fig~\ref{fig:Jemn}.
The dotted line shows the result of Ref.~\cite{ys91b}, while the dot-dashed line corresponds to Eqs.~(\ref{eq:Jnp}) and (\ref{eq:SBHY}), where the bare effective masses are used: $m^*_n=m_p^*=m_N$. Clearly, the simplified approach of an angular-independent cross section in Ref.~\cite{ys91b} leads to overestimated $J_{np}$. The solid curves marked 1--5 are calculated in the Brueckner-Hartree-Fock framework employing different nucleon potentials following Ref.\ \cite{Shternin2017JPhCS}. These potentials include two realistic two-body potentials, Argonne v18 (Av18) and CD-Bonn potential, and two models for the effective three-body forces, Urbana IX (UIX) and the microscopic meson-exchange force (TBFmic), see Refs.\ \cite{Shternin2017JPhCS,Baldo2014PhRvC} for details. One indeed observes a strong (up to a factor of 10) difference between the solid curves at higher densities.
In principle, each microscopic potential leads to its own EOS and, as a consequence, a composition of the beta-stable matter. In this sense, the transport properties such as $J_{np}$ need to be computed along with the equation of state. At present, this approach is not practical. In order to be able to calculate $J_{np}$ for an independent EOS, BSk21 in the present case, it was calculated for each on a grid of baryon densities and proton fractions and then interpolated to a desired composition. Notice that, when $J_{np}$ is calculated for a beta-stable matter consistent with the EOS, the discrepancy between the results for different potentials is preserved (see Ref.\ \cite{Shternin2017JPhCS} for thermal conductivity and shear viscosity). 

The difference between the results in Fig.~\ref{fig:Jnp} is large. This represents the current uncertainty in the microphysical understanding of the properties of the dense NS core matter. Looking at Fig.~\ref{fig:Jnp}, we can recommend to use Eqs.~(\ref{eq:Jnp})--(\ref{eq:SBHY}) with bare nucleon masses (dash-dotted curve) in simulations, bearing in mind at least $\pm 0.5$~dex uncertainty in $J_{np}$ at larger densities. Despite these uncertainties, the $J_{np}$ rate is several orders of magnitude larger than the momentum transfer rates governed by the electromagnetic interactions; see Fig.~\ref{fig:Jemn}.

Finally, we note that the curves 1--5 in Fig.~\ref{fig:Jnp} are calculated under a single many-body approach. Calculations within a different theoretical framework can potentially increase the uncertainty in the rates. For instance, in the medium-modified pion exchange model of nuclear interactions  \cite{Migdal1990PhR}, the pion softening in dense matter can increase the collision cross section by 1--2 orders of magnitude at high densities, which directly translates into the corresponding increase of $J_{np}$. For the shear viscosity, this model is investigated in Ref.~\cite{Kolomeitsev2015PhRvC}.

%%%%%%%%%%%%%%%%%%%%%%%%%%%%%%%%%%%%%%%%%%%%%%%%%%%%%%%%%%%%%%%%%%%%
\section{Comparison with previous works}
\label{sec:toteq}
%%%%%%%%%%%%%%%%%%%%%%%%%%%%%%%%%%%%%%%%%%%%%%%%%%%%%%%%%%%%%%%%%%%%

As the authors believe, 
the formulation of MHD provided in this paper is
simpler than
the analogous formulations 
(e.g., Refs.~\cite{kt08,andersson12,ach17,adhc17,ahdc17})
existing in the literature.
To explain why it is simpler, one should
critically analyze derivations of MHD equations in Refs.\ \cite{kt08,andersson12,adhc17}.
Below we provide such an analysis using, as an example, the recent series of papers \cite{andersson12,adhc17},
which presents the most advanced MHD developed
having in mind applications to NSs.

\begin{enumerate}
	
	\item
The authors of Refs.\ \cite{andersson12,adhc17} obtain a set of evolution equations
	for relativistic magnetized dissipative mixtures,
	having in mind their application to NSs.
	The corresponding equations are derived, separately for each particle species,
	from a variational principle.%
%
%%%%%%
	\footnote{
This approach
proves to be successful
in deriving correct equations
of ordinary and superfluid
hydrodynamics \cite{carter91,prix04,ac07,ac15}.}
%%%%%%%
%
During the derivation the authors do not assume that the relative velocities
	$u^\mu_{(i)}-u^{\mu}_{(j)}$ 
	of different particles species are small.
	The validity of their equations obtained in this way is not discussed;
	they are further used to derive MHD in the approximation
	of small relative velocities.

	A question immediately arises: why should the variational principle provide correct equations
	for a system which is in a highly nonequilibrium kinetic regime?
	It is not clear (at least, for us) what
	is meant by the generalized pressure, chemical potentials, and temperature for the system
	in that case.
	It is well known that, generally, the correct description of such a system
	can be obtained by using kinetic equations for each particle species
	\cite{brag65,abr84}.
	As far as we are aware, nobody has demonstrated that the variational principle used in Refs.\ \cite{andersson12,adhc17}
	is equivalent to the kinetic equation approach.
	\item
	As the second step, the authors analyze the resulting complicated system of equations,
	assuming that the velocity differences $u^\mu_{(i)}-u^{\mu}_{(j)}$ are small (linear drift approximation).
	They also assume that the friction force between different particles
	is proportional to $u^\mu_{(i)}-u^{\mu}_{(j)}$ [see the second term in Eq.~(52) in Ref.~\cite{adhc17}].
	Note that this assumption is not general,
	since in a strong magnetic field the interaction force
	between particles $i$ and $j$ 
	may depend on the magnetic field orientation (e.g., Ref.~\cite{brag65}, Sec.~4).
	Meanwhile, MHD equations developed in the present paper will have the same form
	independently of whether the magnetic field is strong or not:
	effects of magnetic field on particle collisions
	lead only to renormalization of diffusion coefficients
	$\mathcal{D}_{jk}^\parallel$, $\mathcal{D}_{jk}^\perp$, and $\mathcal{D}_{jk}^H$.
	
	\item
	The linearized system of dynamic equations, obtained by the authors 
	[see, e.g., Eqs.~(55) in Ref.~\cite{adhc17}] 
	is still rather complex and is not further simplified (in an essential way).
	We discussed its nonrelativistic analogue in Sec.\ \ref{sec:comparison}
	[see Eq.\ (\ref{eq:ys1})].
	Meanwhile, it is well known (see, e.g., \cite{brag65,og18}, and Sec.\ \ref{sec:comparison})
	that in the hydrodynamic regime ($\mathcal{T} \gg \tau$, $L \gg l$)
	one can significantly simplify these equations
	by replacing the derivatives $\partial_\alpha u^{\mu}_{(j)}$ with $\partial_\alpha u^{\mu}$
	in all equations
	($u^\mu$ can be, e.g., the velocity of center of momentum frame \cite{adhc17}).
	This is exactly the simplification that was made in Sec.~\ref{sec:comparison};
	it allowed us to express the velocities $u^{\mu}_{(j)}$
	algebraically through the gradients of thermodynamic variables
	and establish a connection between our diffusion coefficients
	$\mathcal{D}_{jk}^\parallel$, $\mathcal{D}_{jk}^\perp$, and $\mathcal{D}_{jk}^H$
	and the momentum transfer rates $J_{ik}$.
	
	\item
	Additional complexity of MHD from Refs.\ \cite{andersson12,adhc17}
	is related to
	%is caused by
	numerous entrainment coefficients,
	arising naturally in the variational approach.%
\footnote{In the nondissipative limit, some redundant entrainment coefficients can be set to zero by choosing the appropriate ``gauge'' for the Lagrangian \cite{ga20}.
}
	While the entrainment effect between protons and neutrons 
	is well known in the microscopic theory \cite{bjk96},
	the existence of entrainment between the entropy (treated as a separate fluid)
	and other particle species is not confirmed by microphysics, to the best of our knowledge.%
%
%%%%%%%
\footnote{In Refs.\ \cite{la11,al11}, it is argued that
entropy entrainment is needed to restore causality of the heat conduction equation. }
%%%%%%%
	Setting these coefficients to zero will also simplify the equations.
	
	It is important to note that, although the entrainment between neutrons and protons
	could really affect the dynamic equations written for each particle species separately,
	in our approach, where there is only one velocity field $u^{\mu}$,
	the entrainment does not appear explicitly but leads only to
	renormalization of diffusion coefficients
	$\mathcal{D}_{jk}^\parallel$, $\mathcal{D}_{jk}^\perp$, and $\mathcal{D}_{jk}^H$.
	
    \item
    From a theoretical point of view, the MHD of Refs.\ \cite{andersson12,adhc17} has an advantage that it ensures causality and stability of the resulting equations.
    But how significant are the corresponding corrections to the first-order MHD equations?
    Recent work of Lander and Andersson \cite{la18} 
provides a good example illustrating this point.
These authors analyzed the heat conduction equation in Carter's variational framework. 
They showed that their causal heat equation
reduces to the standard heat equation, used, e.g., in modeling of NS cooling,
if the following condition is satisfied [see their Eq.~(66)]:
\begin{gather}
\label{eq:tau-q}
\tau_Q \gg \frac{\kappa}{c^2 \alpha \hat{s}},
\end{gather}
where $\tau_Q$ is the timescale for variation of the heat flux;
$\kappa$ is the thermal conductivity; 
$\hat{s}$ is the entropy density; $c$ is the speed of light; and $\alpha$ is the lapse function (typically, $\alpha\sim 1$).
Estimating
$\kappa \sim  10^{22}$~erg~cm$^{-1}$~s$^{-1}$~K$^{-1}$ (e.g., \cite{SchmittShternin2018}), $\hat{s}\sim 10^{19}$~erg~cm$^{-3}$~K$^{-1}$ (e.g., \cite{yls99}), one finds that the inequality $\tau_Q \gg \frac{\kappa}{c^2 \alpha  \hat{s}}$ will reduce to $\tau_Q \gg 10^{-18}$~s
(these estimates are made for the temperature $T=10^8$~K).
Since for NS conditions, $\tau_Q\gg 1$~s, it is absolutely safe to use the standard heat equation in all practical situations. 
Note also that, since the thermal conductivity
$\kappa \sim \hat{s} v^2 \tau$  \cite{lp87}, where $v$ is the average microscopic velocity ($v \sim c$ for electrons in NS interiors),
the condition \eqref{eq:tau-q} 
%is similar 
reduces
to $\tau_Q \gg \tau$ (i.e., the system must be in the hydrodynamic regime, which is an expected result).
\end{enumerate}

To sum up, in light of the above discussion it seems quite clear
why the structure of the MHD equations obtained in this paper
appears to be simpler than the formulations available in the literature.

%%%%%%%%%%%%%%%%%%%%%%%%%%%%%%%%%%%%%%%%%%%%%%%%%%%%%%%%%%%%%%%%%%%%
\section{Conclusion}
\label{sec:conclusion}
%%%%%%%%%%%%%%%%%%%%%%%%%%%%%%%%%%%%%%%%%%%%%%%%%%%%%%%%%%%%%%%%%%%%

In the present study we 
formulated equations of dissipative relativistic MHD for nonsuperfluid mixtures.
These equations are rather simple and consist of the energy-momentum conservation law (\ref{eq:dTmunu=0}),
continuity equations for each particle species (\ref{eq:jmu}),
and Maxwell equations (\ref{eq:maxwell-1}) and (\ref{eq:maxwell-2}), 
as well as the second law of thermodynamics (\ref{eq:2ndlaw}).
Dissipative corrections for the particle current densities $\Delta j^{\mu}_{(j)}$
and for the energy-momentum tensor $\Delta\tau^{\mu\nu}$ 
are given by Eqs. (\ref{eq:djmu2}) and (\ref{eq:dtmn2}), respectively.
Ohm's law (\ref{eq:Jfree}) follows automatically from these equations.
Dissipative coefficients, appearing in the proposed MHD,
have a clear physical meaning and can be expressed through
the quantities calculated in the microscopic theory (see Sec.~\ref{sec:comparison}).

How can these MHD equations be used in practice? Let us outline some of the possibilities.
First of all, one can employ the dissipative MHD for studying magnetothermal evolution in the internal layers of NSs,
accounting for diffusion, macroscopic flows, and the effects of general relativity;
the corresponding equations for a spherically symmetric NS with $npe\mu$ core composition were explicitly written out in Sec.~\ref{sec:gr}.
Diffusion may also play an important role in damping of NS oscillations \cite{kg20}, although this effect has not been studied previously, to the best of our knowledge.
As for the development of hydrodynamic theory, the next logical step would be to generalize the dissipative MHD to the superfluid and superconducting mixtures \cite{dg20},
i.e., combine the results of the present paper with the nondissipative superfluid MHD of Ref.\ \cite{gd16}.
This would open up a possibility for realistic modeling of superfluid and superconducting NS cores at finite temperatures.
Finally, let us note that,
while in this paper we were mainly interested in the NS-related applications,
the obtained MHD equations can, in principle, be applied to any relativistic mixture, as long as it stays in the hydrodynamic regime.

%%%%%%%%%%%%%%%%%%%%%%%%%%%%%%%%%%%%%%%%%%%%%%%%%%%%%%%%%%%%%%%%%%%%%%%%%%%%
\subsection*{Acknowledgments}
%%%%%%%%%%%%%%%%%%%%%%%%%%%%%%%%%%%%%%%%%%%%%%%%%%%%%%%%%%%%%%%%%%%%%%%%%%%%

We are very grateful to D.G.~Yakovlev for valuable suggestions and comments.
This study is partially supported by the Foundation 
for the Advancement of Theoretical Physics and Mathematics BASIS 
[Grants  No. 17-12-204-1 (V.A.D. and M.E.G.) and No. 17-13-205-1 (P.S.S.)]
and by Russian Foundation for Basic Research
[Grant No. 19-52-12013].

\appendix

\section{Entropy production due to nonequilibrium reactions}
\label{sec:reaction-rates}

Here
we discuss the entropy production due to nonequilibrium chemical reactions,
which is given by the term
$-\frac{\mu_j}{T} \Delta\Gamma_j$
in Eqs.~\eqref{eq:dSmu-EM2}--\eqref{eq:dSmu-EM-LL}.

Assume for a moment that there is only one chemical reaction between particle species $j=A,B,C,D$:
\begin{gather}
	A + B \leftrightarrow C + D
.
\end{gather}
In this case the source terms $\Delta\Gamma_j$
are constrained by the relations $\Delta\Gamma_A = \Delta\Gamma_B = - \Delta\Gamma_C = - \Delta\Gamma_D \equiv \Delta\Gamma$,
where we introduced the quantity $\Delta\Gamma$.
The reaction is initiated if the chemical potential imbalance
$\Delta\mu \equiv \mu_A + \mu_B - \mu_C - \mu_D$
differs from zero;
thus, in the linear approximation, one can write
\begin{gather}
	\Delta\Gamma = - \lambda \Delta\mu
,	
\end{gather}
where the coefficient $\lambda$ must be positive,
so that the reaction drives the system toward chemical equilibrium.
In terms of $\lambda$ and $\Delta\mu$,
the entropy production rate corresponding to the chemical reaction $A + B \leftrightarrow C + D$
is
\begin{gather}
\label{eq:ldm2}
	-\frac{\mu_j}{T} \Delta\Gamma_j
	= \frac{\lambda}{T} (\mu_A + \mu_B - \mu_C - \mu_D) \Delta \mu
	= \frac{\lambda}{T} (\Delta \mu)^2
.
\end{gather}

Now, let us analyze what happens 
if the reaction takes the form
\begin{gather}
	\alpha_1 A_1 + \alpha_2 A_2 + \cdots
	\leftrightarrow
	\beta_1 B_1 + \beta_2 B_2 + \cdots
,
\end{gather}
where $A_j$ and $B_j$ are particle species; and 
$\alpha_j$ and $\beta_j$ are 
integer stoichiometric coefficients.
In this case one can introduce the reaction rate $\Delta\Gamma$ according to
\begin{gather}
	\Delta\Gamma
	\equiv \frac{\Delta\Gamma_{A_1}}{\alpha_1}
	= \frac{\Delta\Gamma_{A_2}}{\alpha_2}
	= \cdots
	= - \frac{\Delta\Gamma_{B_1}}{\beta_1}
	= - \frac{\Delta\Gamma_{B_2}}{\beta_2}
	= \cdots
,
\end{gather}
and the corresponding conjugate thermodynamic quantity called chemical affinity \cite{KondepoudiPrigogine98} as
\begin{gather}
	\Delta\mu
	\equiv \alpha_1 \mu_{A_1}
		+ \alpha_2 \mu_{A_2}
		+ \cdots
		- \beta_1 \mu_{B_1}
		- \beta_2 \mu_{B_2}
		- \cdots
.
\end{gather}
When all stoichiometric coefficients are equal to $\pm1$, affinity reduces to the chemical potential imbalance used above. With these definitions, one obtains the same result as in the previous case.

Finally, in the most general case of several chemical reactions,
one should proceed by introducing the quantities 
$\lambda_X > 0$, $\Delta\Gamma^X$, and $\Delta\mu_X$
for each reaction $X$,
and, summing up the contribution from all these reactions,
arrive at the final result
\begin{gather}
\label{eq:lambdaXdmuX}
	-\frac{\mu_j}{T} \Delta\Gamma_j
	= \frac{\lambda_X}{T} (\Delta \mu_X)^2
.
\end{gather}

In the case of $npe\mu$ matter
with beta processes turned on
Eq.~\eqref{eq:lambdaXdmuX} reduces to
\begin{gather}
	-\frac{\mu_j}{T} \Delta\Gamma_j
	= \frac{\lambda_{e}}{T} (\Delta \mu_{e})^2
	+ \frac{\lambda_{\mu}}{T} (\Delta \mu_{\mu})^2
,
\end{gather}
where
$\Delta \mu_{e} \equiv \mu_{n} - \mu_{p} - \mu_{e}$ and
$\Delta \mu_{\mu} \equiv \mu_{n} - \mu_{p} - \mu_{\mu}$.

\section{General form of kinetic coefficients}
\label{sec:gen-kinetic}

In this Appendix we derive the most general form of kinetic coefficients
under conditions discussed in Sec.~\ref{sec:general:B}
(homogeneous matter, with the only preferred direction in the comoving frame described by the magnetic induction ${\pmb B}$).
The kinetic coefficients in such matter
can be expressed through
$u^\mu$, $\perp^{\mu\nu}$,
$b^{\mu\nu} \equiv \Fperp^{\mu\nu} / \sqrt{B_\alpha B^\alpha}$,
and (scalar) equilibrium thermodynamic quantities.

Let us start with the coefficient $A_{jk}^{\mu\nu}$.
Generally, it has the form
\begin{gather}
\label{eq:Amn}
	A_{jk}^{\mu\nu}
	= a_{1jk} \perp^{\mu\nu}
	+ a_{2jk} u^\mu u^\nu
	+ a_{3jk} b^{\mu\nu}
	+ a_{4jk} b^{\mu\alpha} b^{\nu}_{~\alpha}
.
\end{gather}
Note that the terms containing $u_\alpha b^{\mu\alpha}$
do not appear here, since they vanish identically
in view of Eq.~\eqref{eq:uFperp=0}.
One can also omit the term $a_{2jk} u^\mu u^\nu$,
since $d_{(k)\nu}$ [see Eq.~\eqref{eq:dmu}] is orthogonal to $u^\nu$
and, therefore, this term does not enter
the expression \eqref{eq:dj} for $\Delta j_{(j)}^\mu$.
The expression~\eqref{eq:Amn} satisfies Eq.\ \eqref{eq:uA=0} automatically.
The Onsager principle~\eqref{eq:B:Ajk=Akj}
reads
\begin{gather}
	A_{jk}^{\mu\nu} ({\pmb B})
	= a_{1jk} \perp^{\mu\nu}
		+ a_{3jk} b^{\mu\nu}
		+ a_{4jk} b^{\mu\alpha} b^{\nu}_{~\alpha}
	= A_{kj}^{\nu\mu} (-{\pmb B})
	= a_{1kj} \perp^{\nu\mu}
		- a_{3kj} b^{\nu\mu}
		+ a_{4kj} b^{\nu\alpha} b^{\mu}_{~\alpha}
.
\end{gather}
Since this condition must be true for all
$b^{\mu\nu}$,
one can conclude that
\begin{gather}
	a_{1jk} = a_{1kj}
,\quad
	a_{3jk} = a_{3kj}
,\quad
	a_{4jk} = a_{4kj}
.
\end{gather}
Using the identity
\begin{gather}
\label{eq:app:bb=perp-bb}
	b^{\mu\alpha} b^{\nu}_{~\alpha}
	= \perp^{\mu\nu} - b^\mu b^\nu
\end{gather}
and introducing the quantities
$\mathcal{D}_{jk}^\parallel \equiv a_{1jk}$,
$\mathcal{D}_{jk}^\perp \equiv a_{1jk} + a_{4jk}$,
and $\mathcal{D}_{jk}^H \equiv a_{3jk}$,
we finally arrive at the following expression,
\begin{gather}
\label{eq:app:Ajk}
	A_{jk}^{\mu\nu}
	= \mathcal{D}_{jk}^\parallel
		b^\mu b^\nu
	+ \mathcal{D}_{jk}^\perp
		\left( \perp^{\mu\nu} - b^\mu b^\nu \right)		
	+ \mathcal{D}_{jk}^H
		b^{\mu\nu}
.
\end{gather}
Similarly, one can show that in the absence of a magnetic field
\begin{gather}
\label{eq:app:Ajk-noB}
	A_{jk}^{\mu\nu}
	= \mathcal{D}_{jk} \perp^{\mu\nu} 
.
\end{gather}

Now, let us consider the coefficients $B^{\mu\nu\lambda}_j$ and $C^{\mu\nu\lambda}_j$.
The general form for $C^{\mu\nu\lambda}_j$, satisfying the constraint \eqref{eq:Cmnl=Cnml},
is
\begin{equation}
\label{eq:app:Cj}
\begin{split}
	C^{\mu\nu\lambda}_j
	= &~c_{1j} u^\mu u^\nu u^\lambda
	+ c_{2j} \left(
		u^\mu \perp^{\nu\lambda}
	  + u^\nu \perp^{\mu\lambda}
		\right)
	+ c_{3j} u^\lambda \perp^{\mu\nu}
	\\
	&+ c_{4j} \left(
		u^\mu b^{\nu\lambda}
	  + u^\nu b^{\mu\lambda}
	\right)
	+ c_{5j} \left(
		u^\mu b^{\nu\alpha} b^{\lambda}_{~\alpha}
	  + u^\nu b^{\mu\alpha} b^{\lambda}_{~\alpha}
	\right)
	+ c_{6j} u^\lambda b^{\mu\alpha} b^{\nu}_{~\alpha}
.
\end{split}
\end{equation}
The first term can be omitted
since it does not enter the expressions for
$\Delta j_{(j)}^\mu$~\eqref{eq:dj}
and for $\Delta\tau^{\mu\nu}$~\eqref{eq:dtau}
in view of the equalities
$u^\mu d_{(j)\mu} = u^\mu \nablaperp_\mu u_\nu = 0$.
Substituting Eq.~\eqref{eq:app:Cj} into the condition~\eqref{eq:uC=0},
one obtains
\begin{gather}
	u_\mu C^{\mu\nu\lambda}_j d_{(j)\lambda}
	= - c_{2j} d_{(j)}^\nu
	  - c_{4j} b^{\nu\lambda} d_{(j)\lambda}
	- c_{5j} b^{\nu\alpha} b^{\lambda}_{~\alpha} d_{(j)\lambda}
	= 0
,
\end{gather}
which implies
\begin{gather}
	c_{2j} = c_{4j} = c_{5j} = 0
.
\end{gather}
Thus, $C^{\mu\nu\lambda}_j$ takes the form
\begin{equation}
\label{eq:app:Cj-2}
	C^{\mu\nu\lambda}_j
	= c_{3j} u^\lambda \perp^{\mu\nu}
	+ c_{6j} u^\lambda b^{\mu\alpha} b^{\nu}_{~\alpha}
.
\end{equation}
Now let us make use of the constraint~\eqref{eq:uB=0} on $B^{\mu\nu\lambda}_j$.
Substituting Eqs.~\eqref{eq:B:C=B} and \eqref{eq:app:Cj-2} into the condition~\eqref{eq:uB=0},
one finds
\begin{gather}
	u_\mu B_{k}^{\mu\nu\lambda} \nablaperp_\nu u_\lambda
	=  c_{3j} \nablaperp_\lambda u^\lambda
	  + c_{6j} b^{\nu\alpha} b^{\lambda}_{~\alpha}
	  		\nablaperp_\nu u_\lambda
	= 0
,
\end{gather}
which can be satisfied for arbitrary
$\nablaperp_\nu u_\lambda$ and $b^{\mu\nu}$
only if $c_{3j} = c_{6j} = 0$.
As a result, we proved that
\begin{gather}
	B^{\mu\nu\lambda}_j = C^{\mu\nu\lambda}_j = 0
.
\end{gather}

Finally, let us consider the tensor $D^{\mu\nu\lambda\sigma}$.
Generally, it can be expressed in terms of the following rank-$4$ tensors:
$u^\mu u^\nu u^\lambda u^\sigma$,
$u^\mu u^\nu \perp^{\lambda\sigma}$,
$u^\mu u^\nu b^{\lambda\sigma}$,
$u^\mu u^\nu b^{\lambda\alpha} b^{\sigma}_{~\alpha}$,
$g^{\mu\nu} \perp^{\lambda\sigma}$,
$g^{\mu\nu} b^{\lambda\sigma}$,
$g^{\mu\nu} b^{\lambda\alpha} b^{\sigma}_{~\alpha}$,
$b^{\mu\nu} b^{\lambda\sigma}$,
$b^{\mu\nu} b^{\lambda\alpha} b^{\sigma}_{~\alpha}$,
and
$b^{\mu\alpha} b^{\nu}_{~\alpha} b^{\lambda\beta} b^{\sigma}_{~\beta}$.
For further convenience, we introduce the tensor
$\Xi^{\mu\nu} \equiv b^{\mu\alpha} b^{\nu}_{~\alpha}$
and, noting that $\perp^{\mu\nu} = \Xi^{\mu\nu} + b^\mu b^\nu$
[see Eq.~\eqref{eq:app:bb=perp-bb}],
express $D^{\mu\nu\lambda\sigma}$
in terms of 
$u^\mu u^\nu u^\lambda u^\sigma$,
$u^\mu u^\nu \Xi^{\lambda\sigma}$,
$u^\mu u^\nu b^{\lambda\sigma}$,
$u^\mu u^\nu b^{\lambda} b^{\sigma}$,
$\Xi^{\mu\nu} \Xi^{\lambda\sigma}$,
$\Xi^{\mu\nu} b^{\lambda\sigma}$,
$\Xi^{\mu\nu} b^{\lambda} b^{\sigma}$,
$b^{\mu\nu} b^{\lambda\sigma}$,
$b^{\mu\nu} b^{\lambda} b^{\sigma}$,
and
$b^{\mu} b^{\nu} b^{\lambda} b^{\sigma}$.
The tensors
$u^\mu u^\nu$,
$\Xi^{\mu\nu}$,
$b^{\mu\nu}$,
and $b^\mu b^\nu$
in the comoving frame, in which
the magnetic field is directed along the $z$ axis, $B=B_z$,
have the following form:
\begin{gather}
	u^\mu u^\nu
	=
	\left(
	\begin{array}{cccc}
		1 & 0 & 0 & 0
		\\
		0 & 0 & 0 & 0
		\\
		0 & 0 & 0 & 0
		\\
		0 & 0 & 0 & 0
	\end{array}
	\right)
,\quad
	\Xi^{\mu\nu}
	=
	\left(
	\begin{array}{cccc}
		0 & 0 & 0 & 0
		\\
		0 & 1 & 0 & 0
		\\
		0 & 0 & 1 & 0
		\\
		0 & 0 & 0 & 0
	\end{array}
	\right)
,\quad
	b^{\mu\nu}
	=
	\left(
	\begin{array}{cccc}
		0 & 0 & 0 & 0
		\\
		0 & 0 & 1 & 0
		\\
		0 & -1 & 0 & 0
		\\
		0 & 0 & 0 & 0
	\end{array}
	\right)
,\quad
	b^\mu b^\nu
	=
	\left(
	\begin{array}{cccc}
		0 & 0 & 0 & 0
		\\
		0 & 0 & 0 & 0
		\\
		0 & 0 & 0 & 0
		\\
		0 & 0 & 0 & 1
	\end{array}
	\right)
.
\end{gather}

The general form of $D^{\mu\nu\lambda\sigma}$,
satisfying conditions \eqref{eq:Dmnls=Dnmls} and \eqref{eq:B:Dmnls=Dlsmn},
reads
\begin{gather}
\label{eq:app:D-1}
\begin{split}
	D^{\mu\nu\lambda\sigma}
	= &~d_1 u^\mu u^\nu u^\lambda u^\sigma
	\\&+ d_2 \left[
			u^\mu u^\nu \Xi^{\lambda\sigma}
			+ u^\lambda u^\sigma \Xi^{\mu\nu}
		\right]
	\\&+ d_3 \left[
			  u^\mu u^\lambda \Xi^{\nu\sigma}
			+ u^\mu u^\sigma \Xi^{\nu\lambda}
			+ u^\nu u^\lambda \Xi^{\mu\sigma}
			+ u^\nu u^\sigma \Xi^{\mu\lambda}
		\right]
	\\&+ d_4 \left[
			u^\mu u^\nu b^\lambda b^\sigma
			+ u^\lambda u^\sigma b^\mu b^\nu
		\right]
	\\&+ d_5 \left[
			u^\mu u^\lambda b^\nu b^\sigma
			+ u^\mu u^\sigma b^\nu b^\lambda
			+ u^\nu u^\lambda b^\mu b^\sigma
			+ u^\nu u^\sigma b^\mu b^\lambda
		\right]
	\\&+ d_6~\Xi^{\mu\nu} \Xi^{\lambda\sigma} 
	\\&+ d_7 \left[
			\Xi^{\mu\lambda} \Xi^{\nu\sigma} 
			+ \Xi^{\mu\sigma} \Xi^{\nu\lambda}
		\right]
	\\&+ d_8 \left[
			\Xi^{\mu\nu} b^\lambda b^\sigma
			+ \Xi^{\lambda\sigma} b^\mu b^\nu
		\right]
	\\&+ d_9 \left[
			\Xi^{\mu\lambda} b^\nu b^\sigma
			+ \Xi^{\mu\sigma} b^\nu b^\lambda
			+ \Xi^{\nu\lambda} b^\mu b^\sigma
			+ \Xi^{\nu\sigma} b^\mu b^\lambda
		\right]
	\\&+ d_{10} ~ b^\mu b^\nu b^\lambda b^\sigma
	\\&+ d_{11} \left[
			u^\mu b^\lambda b^{\nu\sigma}
			+ u^\mu b^\sigma b^{\nu\lambda}
			+ u^\nu b^\lambda b^{\mu\sigma}
			+ u^\nu b^\sigma b^{\mu\lambda}
		\right]
	\\&+ d_{12} \left[
			\Xi^{\mu\lambda} b^{\nu\sigma}
			+ \Xi^{\mu\sigma} b^{\nu\lambda}
			+ \Xi^{\nu\lambda} b^{\mu\sigma}
			+ \Xi^{\nu\sigma} b^{\mu\lambda}
		\right]
	\\&+ d_{13} \left[
			b^\mu b^\lambda b^{\nu\sigma}
			+ b^\mu b^\sigma b^{\nu\lambda}
			+ b^\nu b^\lambda b^{\mu\sigma}
			+ b^\nu b^\sigma b^{\mu\lambda}
		\right]
.
\end{split}
\end{gather}
Here we do not write the term proportional to
$b^{\mu\lambda} b^{\nu\sigma} + b^{\mu\sigma} b^{\nu\lambda}$,
since it can be expressed through other terms~\cite{hsr11}.

Now, let us note that we should omit 
all the terms depending on $u^\lambda$ and $u^\sigma$
in the expression \eqref{eq:app:D-1}
[since
$u^\lambda \nablaperp_\lambda u_\sigma = u^\sigma \nablaperp_\lambda u_\sigma = 0$, 
they do not enter the expression (\ref{eq:dtau}) for $\Delta \tau^{\mu\nu}$]
and also the terms depending on $u^\mu$ and $u^\nu$
[otherwise the constraint~(\ref{eq:uD=0}) is not satisfied].
As a result,
$D^{\mu\nu\lambda\sigma}$ takes the form
\begin{gather}
\begin{split}
	D^{\mu\nu\lambda\sigma}
	= &~d_6~\Xi^{\mu\nu} \Xi^{\lambda\sigma} 
	\\&+ d_7 \left[
			\Xi^{\mu\lambda} \Xi^{\nu\sigma} 
			+ \Xi^{\mu\sigma} \Xi^{\nu\lambda}
		\right]
	\\&+ d_8 \left[
			\Xi^{\mu\nu} b^\lambda b^\sigma
			+ \Xi^{\lambda\sigma} b^\mu b^\nu
		\right]
	\\&+ d_9 \left[
			\Xi^{\mu\lambda} b^\nu b^\sigma
			+ \Xi^{\mu\sigma} b^\nu b^\lambda
			+ \Xi^{\nu\lambda} b^\mu b^\sigma
			+ \Xi^{\nu\sigma} b^\mu b^\lambda
		\right]
	\\&+ d_{10} ~ b^\mu b^\nu b^\lambda b^\sigma
	\\&+ d_{12} \left[
			\Xi^{\mu\lambda} b^{\nu\sigma}
			+ \Xi^{\mu\sigma} b^{\nu\lambda}
			+ \Xi^{\nu\lambda} b^{\mu\sigma}
			+ \Xi^{\nu\sigma} b^{\mu\lambda}
		\right]
	\\&+ d_{13} \left[
			b^\mu b^\lambda b^{\nu\sigma}
			+ b^\mu b^\sigma b^{\nu\lambda}
			+ b^\nu b^\lambda b^{\mu\sigma}
			+ b^\nu b^\sigma b^{\mu\lambda}
		\right]
.
\end{split}
\end{gather}
It has seven independent terms
and, hence, seven viscosity coefficients.
In terms
of viscosity coefficients
introduced in Ref.~\cite{lp87}
$D^{\mu\nu\lambda\sigma}$ can be presented as
\begin{gather}
\label{eq:app:Dmlns-final}
\begin{split}
	D^{\mu\nu\lambda\sigma}
	= &\left( \frac{1}{3}\eta_0 - \eta_1 + \zeta \right) 
			\Xi^{\mu\nu} \Xi^{\lambda\sigma} 
	\\&+ \eta_1 \left[
			\Xi^{\mu\lambda} \Xi^{\nu\sigma} 
			+ \Xi^{\mu\sigma} \Xi^{\nu\lambda}
		\right]
	\\&+ \left( -\frac{2}{3}\eta_0  + \zeta + \zeta_1 \right) \left[
			\Xi^{\mu\nu} b^\lambda b^\sigma
			+ \Xi^{\lambda\sigma} b^\mu b^\nu
		\right]
	\\&+ \eta_2 \left[
			\Xi^{\mu\lambda} b^\nu b^\sigma
			+ \Xi^{\mu\sigma} b^\nu b^\lambda
			+ \Xi^{\nu\lambda} b^\mu b^\sigma
			+ \Xi^{\nu\sigma} b^\mu b^\lambda
		\right]
	\\&+ \left( \frac{4}{3}\eta_0 + \zeta + 2\zeta_1\right)
			b^\mu b^\nu b^\lambda b^\sigma
	\\&+ \frac{1}{2} \eta_3 \left[
			\Xi^{\mu\lambda} b^{\nu\sigma}
			+ \Xi^{\mu\sigma} b^{\nu\lambda}
			+ \Xi^{\nu\lambda} b^{\mu\sigma}
			+ \Xi^{\nu\sigma} b^{\mu\lambda}
		\right]
	\\&+ \eta_4 \left[
			b^\mu b^\lambda b^{\nu\sigma}
			+ b^\mu b^\sigma b^{\nu\lambda}
			+ b^\nu b^\lambda b^{\mu\sigma}
			+ b^\nu b^\sigma b^{\mu\lambda}
		\right]
,
\end{split}
\end{gather}
where
$\eta_0, \ldots \eta_4$ are five shear viscosity coefficients
and $\zeta$ and $\zeta_1$ are two bulk viscosity coefficients,
determining the trace of the tensor $\Delta\tau^{\mu\nu}$ \eqref{eq:dtmn2}:
\begin{gather}
	\Delta\tau^{\mu}_{~\mu}
	= -D_{\mu}^{~\mu\lambda\sigma} \nablaperp_\lambda u_\sigma
	= -\left( 3\zeta + \zeta_1 \right) \Xi^{\lambda\sigma} \nablaperp_\lambda u_\sigma
	- \left( 3\zeta +4\zeta_1 \right) b^\lambda b^\sigma \nablaperp_\lambda u_\sigma
.
\end{gather}
The coefficients $\eta_0$, $\eta_1$, $\eta_2$, $\zeta$, and $\zeta_1$ must be non-negative
in order to ensure entropy growth.
The terms depending on $\eta_3$ and $\eta_4$
do not contribute to the entropy production, so these two coefficients can have arbitrary signs.
The shear viscosity coefficients $\eta_0\dots\eta_4$ have not been calculated for magnetized neutron star cores to our knowledge.
For a magnetized NS crust they were calculated in Ref.~\cite{oy15}
(note that $\eta_3$ and $\eta_4$ in that paper have the opposite sign as compared to our definition).

A similar expression for $D^{\mu\nu\lambda\sigma}$ was derived in Ref.~\cite{hsr11},
where relativistic dissipative MHD of a one-component liquid was analyzed;
note, however, that the authors of Ref.~\cite{hsr11} used
different definitions of viscosity coefficients.

One can show that, in the absence of a magnetic field, 
$D^{\mu\nu\lambda\sigma}$
takes the form \cite{ll87}
\begin{gather}
\label{eq:app:Dmnls-noB}
	D^{\mu\nu\lambda\sigma}
	= \eta \left( \perp^{\mu\lambda} \perp^{\nu\sigma}
			+ \perp^{\nu\lambda} \perp^{\mu\sigma}
			\right)
	+ \left(\zeta - \frac{2}{3} \eta \right)	
		\perp^{\mu\nu} \perp^{\lambda\sigma}
\end{gather}
in order to satisfy the conditions \eqref{eq:Dmnls=Dlsmn} and \eqref{eq:Dmnls=Dnmls}.
Here, $\eta$ and $\zeta$ are (non-negative) shear and bulk viscosity coefficients, respectively.

\section{Diffusion coefficients for N-component plasma}
\label{sec:ncomp}
In this Appendix, we describe how to express diffusion coefficients
$\mathcal{D}_{jk}^\parallel$, $\mathcal{D}_{jk}^\perp$, and $\mathcal{D}_{jk}^H$
through the momentum transfer rates $J_{jk}$
for the case of $N$-component plasma in the low-temperature limit,
considered in Sec.~\ref{sec:comparison}.
To find these relations, we rewrite the system \eqref{eq:mnu=0} and \eqref{eq:ys4}
in terms of particle current perturbations $\Delta {\pmb j}_{(j)} = n_j {\pmb u}_j /c$
and vectors ${\pmb d}_{(j)} = \frac{{\pmb \nabla}\mu_j -e_j {\pmb E}}{T}$,
solve it with respect to $\Delta {\pmb j}_{(j)}$,
and compare the result with Eq.~\eqref{eq:dj-comoving2}.\footnote{
Note that our definitions for 
$\mathcal{D}_{jk}^\parallel$, $\mathcal{D}_{jk}^\perp$, $\mathcal{D}_{jk}^H$,
and ${\pmb d}_{(j)}$
do not coincide with that of Ref.~\cite{ys91a} (see Appendix A there).
}
Here, as in Sec.~\ref{sec:comparison}, we do not set $c = 1$ and do not assume summation over repeated indices.

Being expressed in terms of $\Delta {\pmb j}_{(j)}$ and ${\pmb d}_{(j)}$,
Eqs.\ \eqref{eq:mnu=0} and \eqref{eq:ys4} (divided by $\mu_{j} n_{j}$) read, respectively,
\begin{gather}
\label{eq:ncomp:euler}
	\frac{1}{c^2}
		\frac{\partial {\pmb U}}{\partial t}
	= 
	- \frac{T}{\mu_j} {\pmb d}_{(j)}
	+ \frac{e_j}{\mu_j n_j}\left[ \Delta{\pmb j}_{(j)} \times {\pmb B} \right]
	- \frac{c}{\mu_j n_j}
		\sum_{k \neq j} 
			J_{jk}
			\left(
				\frac{\Delta{\pmb j}_{(j)}}{n_j} - \frac{\Delta{\pmb j}_{(k)}}{n_k}
			\right)
,\\
\label{eq:ncomp:mnu=0}
	\sum_{j} \mu_j \Delta {\pmb j}_{(j)} = 0
.
\end{gather}
Substituting $\Delta {\pmb j}_{(j)}$ from Eq.\ \eqref{eq:dj-comoving2}
into the system~\eqref{eq:ncomp:euler} and \eqref{eq:ncomp:mnu=0}
and taking into account that vectors ${\pmb d}_{(k)\parallel}$ and ${\pmb d}_{(k)\perp}$
are arbitrary, one can express the coefficients
$\mathcal{D}_{jk}^{\parallel}$, $\mathcal{D}_{jk}^{\perp}$, and $\mathcal{D}_{jk}^{H}$
through the momentum transfer rates $J_{jk}$.

Let us present a step-by-step algorithm for finding the coefficients
$\mathcal{D}_{j1}^{\parallel}$, $\mathcal{D}_{j1}^{\perp}$, 
and $\mathcal{D}_{j1}^{H}$ (all other coefficients can be determined in a similar way).
It is sufficient to consider the case when ${\pmb d}_{(k)\parallel} = 0$ and ${\pmb d}_{(k)\perp} = 0$
for all $k \neq 1$.
Then Eq.\ \eqref{eq:ncomp:mnu=0} reduces to
\begin{gather}
	\sum_{j} \mu_j 
	\left( \mathcal{D}_{j1}^{\parallel} {\pmb d}_{(1)\parallel}
		+ \mathcal{D}_{j1}^{\perp} {\pmb d}_{(1)\perp}
		+ \mathcal{D}_{j1}^{H}
			\left[ {\pmb d}_{(1)\perp} \times {\pmb b} \right]
	\right)
	= 0
,
\end{gather}
which implies
\begin{gather}
\label{eq:ncomp:mua=0}
	\sum_{j} \mu_j \mathcal{D}_{j1}^{\parallel} = 0
,\quad
	\sum_{j} \mu_j \mathcal{D}_{j1}^{\perp} = 0
,\quad
	\sum_{j} \mu_j \mathcal{D}_{j1}^{H} = 0
.
\end{gather}
Equation~\eqref{eq:ncomp:euler},
after substituting 
$\Delta {\pmb j}_{(j)}$ from Eq.\ \eqref{eq:dj-comoving2}
and gathering coefficients at
${\pmb d}_{(1)\parallel}$,
${\pmb d}_{(1)\perp}$, and
$\left[ {\pmb d}_{(1)\perp} \times {\pmb b} \right]$,
yields
\begin{gather}
\begin{split}
\label{eq:ncomp:euler-5}
	\frac{1}{c^2}
		\frac{\partial {\pmb U}}{\partial t}
	= 
	&\left[
		- \frac{T}{\mu_j} \delta_{1j}
		- \frac{c}{\mu_j n_j}
			\sum_{k \neq j} 
				J_{jk}
				\left(
					- \frac{\mathcal{D}_{j1}^{\parallel}}{n_j}
					+ \frac{\mathcal{D}_{k1}^{\parallel}}{n_k}
				\right)
	\right]
	{\pmb d}_{(1)\parallel}
\\
	&+
	\left[
		- \frac{T}{\mu_j} \delta_{1j}
		+ \frac{e_j B}{\mu_j n_j}
			\mathcal{D}_{j1}^{H}
		- \frac{c}{\mu_j n_j}
			\sum_{k \neq j} 
				J_{jk}
				\left(
					- \frac{\mathcal{D}_{j1}^{\perp}}{n_j}
					+ \frac{\mathcal{D}_{k1}^{\perp}}{n_k}
				\right)
	\right]
	{\pmb d}_{(1)\perp}	
\\
	&+
	\left[
		- \frac{e_j B}{\mu_j n_j}
			\mathcal{D}_{j1}^{\perp}
		- \frac{c}{\mu_j n_j}
			\sum_{k \neq j} 
				J_{jk}
				\left(
					- \frac{\mathcal{D}_{j1}^{H}}{n_j}
					+ \frac{\mathcal{D}_{k1}^{H}}{n_k}
				\right)
	\right]
	\left[ {\pmb d}_{(1)\perp} \times {\pmb b} \right]
.
\end{split}
\end{gather}

Now let us subtract Eq.~\eqref{eq:ncomp:euler} for $j=2,3,\ldots N$
from Eq.~\eqref{eq:ncomp:euler} for $j = 1$.
Equating  coefficients at
${\pmb d}_{(1)\parallel}$,
${\pmb d}_{(1)\perp}$, and
$\left[ {\pmb d}_{(1)\perp} \times {\pmb b} \right]$ to zero,
we obtain
\begin{gather}
\label{eq:ncomp:euler-dpar}
	\left[
		- \frac{c}{\mu_1 n_1}
			\sum_{k \neq 1} 
				J_{1k}
				\left(
					- \frac{\mathcal{D}_{11}^{\parallel}}{n_1}
					+ \frac{\mathcal{D}_{k1}^{\parallel}}{n_k}
				\right)
	\right]
	- \left[
			- \frac{c}{\mu_j n_j}
				\sum_{k \neq j} 
					J_{jk}
					\left(
						- \frac{\mathcal{D}_{j1}^{\parallel}}{n_j}
						+ \frac{\mathcal{D}_{k1}^{\parallel}}{n_k}
					\right)
		\right]
	= \frac{T}{\mu_1}
,\quad j = 2 \ldots N
,\\
\label{eq:ncomp:euler-dperp}
	\left[
		\frac{e_1 B}{\mu_1 n_1}
			\mathcal{D}_{11}^{H}
		- \frac{c}{\mu_1 n_1}
			\sum_{k \neq 1} 
				J_{1k}
				\left(
					- \frac{\mathcal{D}_{11}^{\perp}}{n_1}
					+ \frac{\mathcal{D}_{k1}^{\perp}}{n_k}
				\right)
	\right]
	- \left[
		\frac{e_j B}{\mu_j n_j}
			\mathcal{D}_{j1}^{H}
		- \frac{c}{\mu_j n_j}
			\sum_{k \neq j} 
				J_{jk}
				\left(
					- \frac{\mathcal{D}_{j1}^{\perp}}{n_j}
					+ \frac{\mathcal{D}_{k1}^{\perp}}{n_k}
				\right)
	\right]
	= \frac{T}{\mu_1}
,\quad j = 2 \ldots N
,\\
\label{eq:ncomp:euler-dB}
	\left[
		- \frac{e_1}{\mu_1 n_1}
			\mathcal{D}_{11}^{\perp}
		- \frac{c}{\mu_1 n_1}
			\sum_{k \neq 1} 
				J_{1k}
				\left(
					- \frac{\mathcal{D}_{11}^{H}}{n_1}
					+ \frac{\mathcal{D}_{k1}^{H}}{n_k}
				\right)
	\right]
	-
	\left[
		- \frac{e_j}{\mu_j n_j}
			\mathcal{D}_{j1}^{\perp}
		- \frac{c}{\mu_j n_j}
			\sum_{k \neq j} 
				J_{jk}
				\left(
					- \frac{\mathcal{D}_{j1}^{H}}{n_j}
					+ \frac{\mathcal{D}_{k1}^{H}}{n_k}
				\right)
	\right]
	= 0
,\quad j = 2 \ldots N
.
\end{gather}
Equations~\eqref{eq:ncomp:euler-dpar}--\eqref{eq:ncomp:euler-dB} for $j=2,\ldots N$
together with the conditions \eqref{eq:ncomp:mua=0}
can be written, in the matrix form, as
\begin{gather}
\label{eq:ncomp:matrixform}
\left[
\begin{array}{c|c|c}
	\mathcal{M}^{\parallel(1)} & 0 & 0
	\\ \hline
	0 & \mathcal{M}^{\parallel(1)} & \mathcal{M}^{H(1)}
	\\ \hline
	0 & - \mathcal{M}^{H(1)} & \mathcal{M}^{\parallel(1)}
\end{array}
\right]
\left[
\begin{array}{c}
	\mathit{X}^{\parallel(1)}
	\\ \hline
	\mathit{X}^{\perp(1)}
	\\ \hline
	\mathit{X}^{H(1)}
\end{array}
\right]
=
\left[
\begin{array}{c}
	\mathit{Y}^{(1)}
	\\ \hline
	\mathit{Y}^{(1)}
	\\ \hline
	0
\end{array}
\right]
,
\end{gather}
where $\mathit{X}^{\parallel(1)}$,
$\mathit{X}^{\perp(1)}$,
$\mathit{X}^{H(1)}$,
and $\mathit{Y}^{(1)}$
are $N$-dimensional vectors,
\begin{gather}
	\mathit{X}^{\parallel(1)} = (\mathcal{D}_{11}^{\parallel}, \mathcal{D}_{21}^{\parallel}, \ldots \mathcal{D}_{N1}^{\parallel})^T
,\\
	\mathit{X}^{\perp(1)} = (\mathcal{D}_{11}^{\perp}, \mathcal{D}_{21}^{\perp}, \ldots \mathcal{D}_{N1}^{\perp})^T
,\\
	\mathit{X}^{H(1)} = (\mathcal{D}_{11}^{H}, \mathcal{D}_{21}^{H}, \ldots \mathcal{D}_{N1}^{H})^T
,\\
	\mathit{Y}^{(1)} = \left( 0, \frac{T}{\mu_1}, \frac{T}{\mu_1},\ldots \frac{T}{\mu_1} \right)^T
,
\end{gather}
while
$\mathcal{M}^{\parallel(1)}$ 
and $\mathcal{M}^{H(1)}$
are $N \times N$ matrices, whose elements are determined as
($\delta_{jk}$ is the Kronecker delta)
\begin{gather}
	\mathcal{M}^{\parallel(1)}_{1k} = \mu_k
,\quad
	k = 1 \ldots N
,\\
	\mathcal{M}^{\parallel(1)}_{j1}
	= \frac{c}{\mu_1 n_1} \frac{\sum_{l\neq 1} J_{1l}}{n_1}
	+ \frac{c}{\mu_j n_j} \frac{J_{j1}}{n_1}
,\quad
	j = 2 \ldots N	
,\\
	\mathcal{M}^{\parallel(1)}_{jk}
	= - \frac{c}{\mu_1 n_1} \frac{J_{1k}}{n_k}
	  + \frac{c}{\mu_j n_j} \frac{J_{jk}}{n_k}
	  - \delta_{jk}
	  	  \frac{c}{\mu_j n_j} \frac{\sum_{l\neq j} J_{jl}}{n_j}
,\quad
	j,k = 2 \ldots N
,\\
	\mathcal{M}^{H(1)}_{jk}
	= \frac{e_1 B}{\mu_1 n_1} \delta_{1k}
	- \frac{e_j B}{\mu_j n_j} \delta_{jk}
,\quad
	j,k = 1 \ldots N.
\end{gather}

Solving Eq.~\eqref{eq:ncomp:matrixform},
one can find the coefficients $\mathcal{D}_{j1}^{\parallel}$, $\mathcal{D}_{j1}^{\perp}$, and $\mathcal{D}_{j1}^{H}$.
Matrices
$\mathcal{D}_{jk}^{\parallel}$, $\mathcal{D}_{jk}^{\perp}$, and $\mathcal{D}_{jk}^{H}$ with $k\neq 1$
can be obtained in a similar way.
One can also directly check that these matrices are symmetric; i.e., the Onsager principle is satisfied
(remember that $J_{jk} = J_{kj}$).

%%%%%%%%%%%%%%%%%%%%%%%%%%%%%%%%%%%%%%%%%%%%%%%%%%%%%%%%%%%%%%%%%%%%%%%%%%%%%%%%%%%%%%
\bibliography{litt}
%%%%%%%%%%%%%%%%%%%%%%%%%%%%%%%%%%%%%%%%%%%%%%%%%%%%%%%%%%%%%%%%%%%%%%%%%%%%%%%%%%%%%

\end{document}